\documentclass[12pt,preprint]{aastex}
\usepackage{epsf}

\newcommand{\sect}[1]{\S\,\ref{#1}}
\newcommand{\be}{\begin{displaymath}}
\newcommand{\ee}{\end{displaymath}}
\newcommand{\bea}{\begin{eqnarray}}
\newcommand{\eea}{\end{eqnarray}}

\shortauthors{Denissenkov, Pinsonneault, \& MacGregor}
\shorttitle{Magneto\,-Thermohaline Mixing in Red Giants}

\begin{document}

\title{MAGNETO\,-THERMOHALINE MIXING IN RED GIANTS}

\author{Pavel A. Denissenkov\altaffilmark{1,2}, Marc Pinsonneault\altaffilmark{1}, and Keith B. MacGregor\altaffilmark{3}}
\altaffiltext{1}{Department of Astronomy, The Ohio State University, 4055 McPherson Laboratory,
       140 West 18th Avenue, Columbus, OH 43210; dpa@astronomy.ohio-state.edu, pinsono@astronomy.ohio-state.edu.}
\altaffiltext{2}{On leave from Sobolev Astronomical Institute of St. Petersburg State University,
   Universitetsky Pr. 28, Petrodvorets, 198504 St. Petersburg, Russia.}
\altaffiltext{3}{High Altitude Observatory, National Center for Atmospheric Research, P.O. Box 3000,
   Boulder, CO 80307-3000; kmac@hao.ucar.edu.}
 
\begin{abstract}
We revise a magnetic buoyancy model that has recently been proposed as a mechanism for
extra mixing in the radiative zones of low-mass red giants. The most important revision is
our accounting of the heat exchange between rising magnetic flux rings and their surrounding
medium. This increases the buoyant rising time by five orders of magnitude, therefore 
the number of magnetic flux rings participating in the mixing has to be increased correspondingly.
On the other hand, our revised model takes advantage of the fact that the mean molecular weight of the rings
formed in the vicinity of the hydrogen burning shell has been reduced by $^3$He burning. This
increases their thermohaline buoyancy (hence, decreases the total ring number) considerably, 
making it equivalent to the pure magnetic buoyancy produced by
a frozen-in toroidal field with $B_\varphi\approx 10$\,MG. We emphasize that some toroidal field
is still needed for the rings to remain cohesive while rising. Besides, this field prevents 
the horizontal turbulent diffusion from eroding the $\mu$ contrast between the rings and their
surrounding medium. We propose that the necessary toroidal magnetic field is generated
by differential rotation of the radiative zone, that stretches a pre-existing poloidal field around the rotation axis,
and that magnetic flux rings are formed as a result of its buoyancy-related instability.
\end{abstract} 


\section{Introduction}

During their first ascent on the red giant branch (RGB),
a majority of low-mass stars (those with $M\la 2\,M_\odot$) experience extra mixing in their radiative zones
separating the H burning shell from the bottom of convective envelope 
(\citealt{sm79,chdn98,dv03}). Despite 30 years of effort, however, the underlying physical mechanism
is still not understood. Observationally, the RGB extra mixing manifests itself through
changes of the surface abundances of Li, C, N, and of the isotopic ratio $^{12}$C/$^{13}$C
correlating with an increasing luminosity (\citealt{grea00,sm03}). 
These changes are produced by the joint operation of thermonuclear reactions that take place 
in the vicinity of the H shell and a nonconvective mixing process that transports reaction products through the radiative zone
to the convective envelope. Observations support the idea that this mixing process starts (or, gets much more efficient)
when an RGB star reaches a luminosity at which the differential luminosity function for a population of stars
having the same age and chemical composition shows a prominent bump (a local pile-up of stars). The luminosity bump
results from a temporary slowing down of the star's evolution caused by its structural readjustment.
This happens when the H shell crosses and erases a discontinuity in the H-abundance profile left behind by
the bottom of convective envelope at the end of the first dredge-up. During the first dredge-up, that occurs on the subgiant branch and on the lower RGB,
the convective envelope grows in mass, which causes its bottom to penetrate the layers whose chemical composition
had been altered yet on the main sequence (MS). This produces changes of the surface abundances of
Li, C, N, and of the $^{12}$C/$^{13}$C ratio similar to but by far less substantial than those 
incurred from the subsequent operation of the RGB extra mixing.

Until recently, it has been thought that the only reason why the RGB extra mixing does not manifest itself
below the bump luminosity is a strong gradient of the mean molecular weight $\mu$ caused by the onset of
a deep convective envelope (e.g., \citealt{chea98}). Any mixing mechanism
has to overcome the stable thermal stratification of the radiative zone, and in the presence of
a positive $\nabla_\mu$ such mixing is correspondingly more difficult. 
In an RGB star above the bump luminosity, the H shell has already crossed the H-profile
discontinuity, therefore the radiative zone is now chemically uniform everywhere except in a very close neighborhood of
the H shell. This circumstance was repeatedly emphasized in the past. 
In particular, it has been used to model the RGB extra mixing with rotation-driven
meridional circulation and turbulent diffusion. It was not until recently that it has become clear that rotational mixing
fails to explain the chemical element transport in the radiative zones of upper RGB stars (\citealt{chea05,pea06}). 
In short, this failure is due to the following main causes: firstly, rotation period measurements for young cluster stars
and helioseismic data indicate that MS stars with $M\la 1\,M_\odot$ loose a great amount of their initial angular momentum via
magnetized stellar winds and that they most likely become slow and nearly solid-body rotators before leaving the MS; 
secondly, the chemical element transport by meridional
circulation is strongly hindered by rotation-induced horizontal turbulence in stellar radiative zones (\citealt{chz92});
thirdly, the vertical turbulent diffusion powered by differential rotation in the radiative zones of RGB stars operates
at a low level too because it also redistributes the angular momentum, thus reducing the degree of differential rotation
in a self-regulating way.

A new class of RGB extra mixing models has emerged since \cite{eea06} noticed that 
a tiny $\mu$-gradient inversion ($\nabla_\mu \approx -10^{-4}$) becomes visible at the outer tail of the H burning shell
precisely at the moment when the H shell erases the H-profile discontinuity.
This inversion is produced by the reaction $^3$He($^3$He,\,2p)$^4$He that locally reduces the mean molecular weight by $\Delta\mu\approx\mu^2\Delta X_3/6$,
where $X_3$ is the $^3$He mass fraction. The mechanism can be effective in the low-mass RGB stars 
because their MS progenitors synthesize large amounts of $^3$He in their outer radiative cores through non-equilibrium pp burning.
Even though this $^3$He-rich material gets diluted in the convective envelope during the first dredge-up,
the radiative zone of a low-mass RGB star above the bump luminosity can still have $X_3$ increased up to a value of $2\times 10^{-3}$
(the solar initial $^3$He abundance is $3\times 10^{-5}$). For this mass fraction,
the $^3$He burning leads to $\Delta\mu\approx -10^{-4}$, assuming that $\mu\approx 0.6$ and $\Delta X_3\approx -X_3$. 
Below the bump luminosity, the $\mu$-gradient inversion is overridden by the strong positive $\mu$-gradient built up
on the MS. It shows up and may come into play only when the $^3$He burning shell, advancing in mass in front of the major H shell, finds itself
in the chemically homogeneous part of the radiative zone. This happens at the bump luminosity.
\cite{eea06} found that a rapid mixing process occurred in their 3D simulations above this point, 
although the underlying cause was not identified (see \citealt{dp08b}).

Inspired by this work, \cite{chz07a} have proposed that the $\mu$-gradient inversion maintained by the $^3$He burning drives thermohaline convection in
the radiative zones of low-mass RGB stars above the bump luminosity and that this is the long-sought physical mechanism for the RGB extra mixing. 
Thermohaline convection is a mixing process triggered by a double diffusive
instability (e.g., \citealt{v04}). Consider a stratified ideal gas with a stable temperature gradient 
($\nabla\equiv d\ln T/d\ln P < \nabla_{\rm ad}$, where ``ad'' stands for adiabatic changes)
but with an unstable composition gradient ($\nabla_\mu < 0$). If we isolate a gas blob and shift it up in the vertical direction
then its further motion will depend on how fast the blob exchanges heat and composition with its surrounding medium horizontally.
Indeed, the relative difference in density between the surrounding medium and the blob is $\Delta\rho/\rho \approx \Delta\mu/\mu - \Delta T/T$,
assuming that $\Delta P = 0$. For the blob to continue rising, we need $\Delta\rho > 0$. Our assumptions about the gradients mean that
$\Delta\mu > 0$ and $\Delta T > 0$ in the absence of both heat and molecular diffusion. Because these differences grow when
the blob rises, $\Delta\rho$ may stay positive or it may ultimately become negative depending on the ratio $r_\mu = |\nabla_\mu|/(\nabla_{\rm ad}-\nabla)$.
In our particular case, $r_\mu\ll 1$. Therefore, our idealized impermeable and adiabatic blob will rise a short distance and then stop, when
the accumulated difference in $T$ compensates that in $\mu$.
In reality, the heat exchange, whose rate is specified by
the radiative diffusivity $K$, constantly works to reduce the difference in $T$. On the other hand, molecular diffusion $\nu_{\rm mol}$
tries to smooth out the difference in $\mu$. The double diffusive instability may therefore develop only if $K\gg \nu_{\rm mol}$.
In this case, the blob's rising speed can be estimated as $v\sim l/\tau_{\rm th}$, where $l$ is the mean path that the blob travels before
it gets dissolved, while $\tau_{\rm th}\sim d^2/K$ is the characteristic thermal time scale for a spherical blob of the diameter $d$.
An approximate expression for the thermohaline diffusion coefficient can be obtained as
$D_{\rm thc}\sim lvr_\mu \sim Kr_\mu(l/d)^2$. \cite{chz07a} and \cite{dp08b} have demonstrated that the observed RGB mixing patterns can be explained by
stellar evolutionary models with the $^3$He-driven thermohaline convection only if $l/d\ga 10$\,--\,$30$. A similarly large parameter ratio for
thermohaline convection in stellar radiative zones was postulated by \cite{u72}, as opposed to a ratio $l/d\sim 1$ advocated by \cite{kea80}.
Besides, the double-diffusive instability has been shown to result in formation of elongated (large $l$ to $d$ ratios) structures
known as ``salt fingers'' in laboratory experiments with the saltier and warmer water overlying the fresher and colder water (\citealt{s60}). 

However, there appears to exist observational and theoretical
arguments challenging this model. First of all, a large number of old metal-poor MS stars with $M\la 0.9\,M_\odot$, both in globular clusters and
in the halo field, that had accreted He- and C-rich high-$\mu$ material from their evolved cluster or binary companions
do not seem to have been thoroughly mixed by thermohaline convection (\citealt{nt07,dp08a,aea08}), as it would be expected even
if the less efficient prescription by \cite{kea80} were used for $D_{\rm thc}$. Second, thermohaline convection is expected to be suppressed
by the rotation-induced horizontal turbulence that works together with the molecular diffusion to reduce the $\mu$ contrast
between the rising gas blob and its surroundings (\citealt{dp08b}). Third, strong differential rotation is predicted to hinder 
thermohaline convection as well, because the ``salt fingers'' may be tilted by the rotational shear so rapidly that they will get damped before
they produce significant mixing (\citealt{c99}, and references therein). We anticipate that a similar effect is also produced by
the Coriolis force in a uniformly rotating radiative zone. Contrary to these expectations, a much larger fraction of
Li-rich objects has been found among rapidly rotating ($v\sin\,i\geq 8$\,km\,s$^{-1}$) K giants than
among their more common slowly rotating ($v\sin\,i\la 1$\,km\,s$^{-1}$) counterparts (\citealt{dea02}). 
The Li-rich K giants are low-mass stars located above the bump luminosity (\citealt{chb00}) in which large amounts of Li are thought to be synthesized      
via the $^7$Be-transport mechanism (\citealt{cf71}). To be efficient, this mechanism needs a 10 to 100 times faster mixing
than that required to reproduce the abundance patterns in the majority of upper RGB stars (\citealt{dh04}). It is not clear how thermohaline convection
can explain the phenomenon of Li-rich K giants given that its efficiency should be lower in the more rapidly rotating stars.
These arguments have motivated our search for an alternative RGB mixing mechanism.

In this paper, we use a simple model of toroidal magnetic field generation in a differentially rotating radiative zone of a bump-luminosity RGB star 
to obtain order-of-magnitude estimates demonstrating that the buoyant rise of magnetic flux rings being
formed close to the local minimum in $\mu$ may be a good alternative to the $^3$He-driven thermohaline convection.
A similar model has recently been proposed by \cite{bea07} (hereafter, referred to as BWNC). However, they assumed that
a rising ring always stays in thermal equilibrium with its surrounding medium, and we argue that this leads to a substantial
overestimate of the ring's radial velocity. We account for the impact of $\mu$ gradients and discuss the origin of
the magnetic rings.  We show that, as a mechanism for the RGB extra mixing,
magnetic buoyancy has some advantages over thermohaline convection
and, therefore, it is worth further investigating by means of multidimensional MHD simulations.

\section{The RGB Stellar Model}

\begin{figure}
\plotone{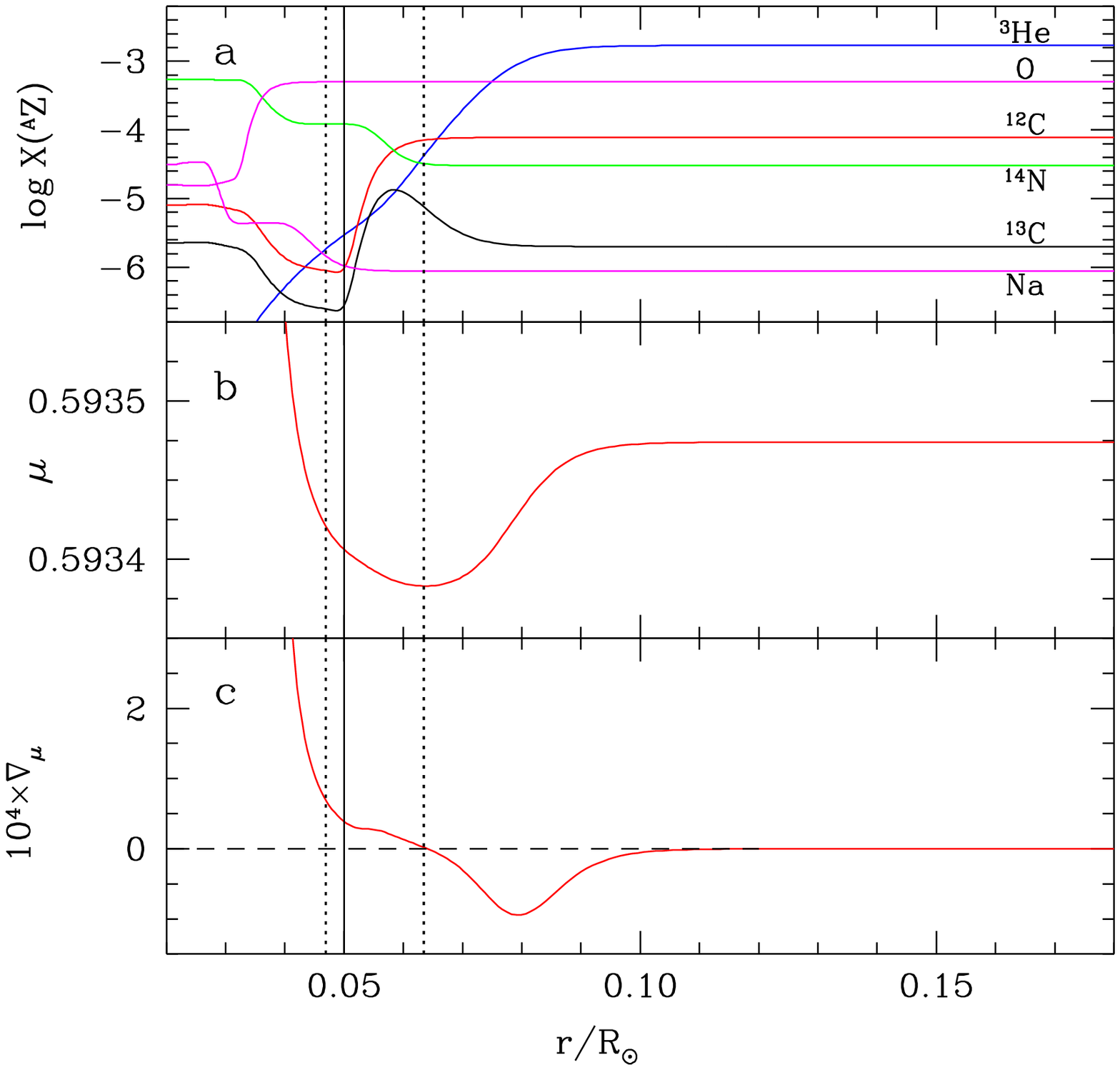}
\caption{(a) The element mass fractions, (b) the mean molecular weight, and (c) its logarithmic (with respect
         to the pressure) gradient as functions of the radius in the vicinity of the H burning shell
         in our bump-luminosity RGB model. The vertical solid line shows the observationally constrained depth of 
         the RGB extra mixing $r_{\rm mix} = 0.05\,R_\odot$, the right dotted line -- the radius $r_{\rm min}$ of
         the minimum $\mu$, while the left dotted line is placed at $r = r_{\rm min} - H_P$.
         }
\label{fig:f1}
\end{figure}

Our background RGB stellar model, in the radiative zone of which the formation and buoyant rise of magnetic flux rings are studied, represents a typical metal-poor upper RGB
star. It has the initial mass $M=0.8\,M_\odot$, helium and heavy-element mass fractions $Y=0.24$ and $Z=0.0005$, and
the luminosity $\log\,L/L_\odot = 2.085$ corresponding to the age of 13.65 Gyrs. The RGB extra mixing is free to work in this star
because the H burning shell has already erased the H-profile discontinuity in it.
The model has been computed using the stellar evolution code described by \cite{dea06}.
Its chemical element mass fraction, $\mu$, and $\nabla_\mu$ profiles immediately above the H shell are plotted in Fig.~\ref{fig:f1}.

For extra mixing to dredge up material deficient in C but not enriched in Na, as required by the observed abundance patterns
in the metal-poor field RGB stars (\citealt{grea00}), it has to reach a depth between $0.045\,R_\odot$ and 
$0.055\,R_\odot$ (Fig.~\ref{fig:f1}a). We will assume the mixing depth
$r_{\rm mix} = 0.05\,R_\odot$ (shown by the vertical solid line in Fig.~\ref{fig:f1}). In previous works (e.g., \citealt{dw96,dv03}),
the mixing depth was specified using either the relative mass coordinate $\delta M_{\rm mix} = (M_{\rm mix}-M_{\rm c})/(M_{\rm bce}-M_{\rm c})$ or
the logarithmic temperature difference $\Delta\log T = \log T(r_{\rm c}) - \log T(r_{\rm mix})$, where the subscripts ``c'' and ``bce'' refer
to the He core boundary and to the bottom of convective envelope, respectively. However, we have noticed that, when plotted as functions of $r$, 
the abundance profiles remain nearly stationary in spite of the slow mass inflow from the radiative zone that feeds the H burning shell.
Therefore, we have decided to simply use the radius for the RGB mixing depth specification. The value of $r_{\rm mix} = 0.05\,R_\odot$
corresponds to $\delta M_{\rm mix} = 0.135$ and $\Delta\log T = 0.245$.

For subsequent estimates of various quantities characterizing the efficiencies of magnetic flux ring formation and buoyant rise in
the radiative zone of our RGB model we need to know some of its structure parameters   
at $r=r_{\rm mix}$. These are summarized in Table~\ref{tab:tab1} along with the parameter values at $r=r_{\rm bce}$.
We have also listed the values used by BWNC. Note that they have considered a half-solar metallicity bump luminosity
model with the initial mass of $1.5\,M_\odot$.

\section{Rotation in the Radiative Zone}

In our magnetic buoyancy model, it is assumed that a toroidal magnetic field in the radiative zone of
an RGB star is generated by its differential rotation that stretches a pre-existing poloidal field around the rotation axis.
To elaborate on the model, we therefore need an estimate of the degree of differential rotation in the radiative zone.
For this, we use the rotation profile M2 presented by \cite{pea06} in their Fig.~2 (the dotted curve in upper panel C). It shows the angular velocity $\Omega$ as
a function of $\delta M = (M_r-M_{\rm c})/(M_{\rm bce}-M_{\rm c})$ in the radiative zone of a bump luminosity model
with initial values of $M$, $Y$, and $Z$ almost identical to ours. The model's rotational
evolution was computed using the stellar evolution code STAREVOL designed for 1D simulations of 
the angular momentum and chemical element transport in stellar radiative zones (\citealt{sea00,pea03}).
The code takes proper account of
the rotation-induced meridional circulation, horizontal and vertical turbulent diffusion, as well as atomic diffusion and mass-loss.
The initial zero-age MS model for the M2 run has been assumed to be a slow solid-body rotator with the surface rotational velocity of 5 km\,s$^{-1}$.
This choice is a reasonable replacement for detailed computations of the pre-MS and early MS evolution during which
a low-mass star experiences a strong magnetic breaking of its much faster initial surface rotation
and a core-envelope rotational coupling leading to its nearly uniform internal rotation on a much shorter time scale than
its MS life time. 

Another important assumption made by \cite{pea06} is uniform specific angular
momentum distribution in the RGB convective envelope. 
This is required to explain the origin of rapidly rotating red horizontal branch stars (\citealt{sp00}).
This empirical demand for a strong differential rotation in the convective envelopes of low-mass RGB stars is supported by recent 3D hydrodynamic
simulations of the interaction of turbulent convection and rotation performed by \cite{pb06}. 

The M2 bump luminosity model has $\Omega(r_{\rm bce})\approx 10^{-6}$ rad\,s$^{-1}$,
and $\Omega(r_{\rm mix})\approx 10^{-3}$ rad\,s$^{-1}$. We will use these values in our following discussion.
So, in the absence of transport processes other than those considered by \cite{pea06}, the angular velocity in the radiative zone 
of our RGB stellar model could vary with the radius
approximately as $\Omega(r) = \Omega(r_{\rm mix})(r_{\rm mix}/r)^2$. This steep rotation profile results from the conservation of
angular momentum in the mass inflow and modest angular momentum redistribution by meridional circulation and vertical turbulent diffusion.

\section{Suppression of Thermohaline Convection by Horizontal Turbulence}
\label{sec:suppress}

\cite{dp08b} have derived a general criterion for convective instability in the presence of a negative $\nabla_\mu$
and strong horizontal turbulence, the latter being  characterized by a diffusion coefficient $D_{\rm h}\gg \nu_{\rm mol}$. They have allowed for a
possibility that a convective element of the diameter $d$ can travel a distance $l > d$. One of the consequences of this criterion is
that thermohaline convection may be suppressed by the horizontal turbulent diffusion unless
\bea
\frac{|\nabla_\mu|}{\nabla_{\rm rad}-\nabla_{\rm ad}} > \frac{1}{3}\frac{D_{\rm h}}{K+D_{\rm h}},
\label{eq:ourcrit}
\eea
where $\nabla_{\rm rad}$ is the radiative temperature gradient
(compare this inequality with condition 5 from \citealt{v04}).
Note that a similar result can be obtained from the dispersion relation (10) derived and analysed by \cite{u72}.
One should only take into account that, in the presence of the strong horizontal turbulence,
Ulrich's molecular diffusion coefficient $\mathbf{D}$ has to be replaced with $D_{\rm h}$. Then, for the thermally
limited modes, one readily finds
\bea
D_{\rm thc} \approx D_{\rm Ulrich}\times \left(1-\frac{D_{\rm h}}{K}\frac{\nabla_{\rm rad}-\nabla_{\rm ad}}{|\nabla_\mu|}\right),
\label{eq:modulrich}
\eea
where $D_{\rm Ulrich}$ is Ulrich's original thermohaline diffusion coefficient.
From the last expression it follows that thermohaline convection may operate ($D_{\rm thc} > 0$) only when
$|\nabla_\mu|/(\nabla_{\rm ad}-\nabla_{\rm rad}) > D_{\rm h}/K$, which is close to the requirement (\ref{eq:ourcrit}),
provided that $D_{\rm h} < K$. The difference in the right-hand-side ratios between the two conditions comes about from the fact that, unlike \cite{dp08b},
Ulrich neglected the contribution of $\mathbf{D}$ to the heat diffusion. That was warranted for the nonrotating case considered by him, in which
$D_{\rm h}=0$, and therefore $\mathbf{D}=\nu_{\rm mol}\ll K$.

\cite{pea06} have shown that $D_{\rm h}$ always stays comparable to $K$ at $r\approx r_{\rm mix}$ in their M2 RGB model. Moreover,
the ratio $D_{\rm h}/K$ turns out to be larger than $\sim$\,$10^{-3}$ at all radii between $r_{\rm mix}$ and $r_{\rm bce}$, both in the bump luminosity
model and in the more evolved model shown in their Fig.~5 (panels C and D). Given that the $^3$He burning can make the ratio 
$|\nabla_\mu|/(\nabla_{\rm rad}-\nabla_{\rm ad}) \sim 10^{-3}$ at most, while the RGB extra mixing will necessarily reduce it
below this limit (by smoothing out the $\mu$-gradient), the linear analysis predicts that thermohaline convection 
may be suppressed in these models, especially in the vicinity of $r_{\rm mix}$. 
It would be inconsistent to ignore this prediction because similar eroding effects of the horizontal turbulence
on the model rotational and mixing properties have already been included in the simulations performed by \cite{pea06}. They are responsible for
the significant reduction of the efficiency of mixing by meridional circulation (\citealt{chz92}) and for the erasing of
the latitudinal differential rotation, the latter effect allowing to consider $\Omega$ as a function of $r$ alone (\citealt{z92}).
Besides, Zahn's concept of the rotation-induced anisotropic turbulence in stellar radiative zones, with horizontal 
components of the turbulent viscosity strongly dominating over those in the vertical direction,
was repeatedly used in stellar evolution computations to facilitate the penetration of $\mu$-gradient barriers by the
vertical turbulent diffusion both in MS stars and in upper RGB stars (\citealt{tz97,tea97,m03,pea03,pea06}).

\section{Generation of Toroidal Magnetic Field}
\label{sec:gen}

Following \cite{mw87}, we assume that, like in the solar-type MS stars, the differential rotation in the radiative zone of the RGB star
creates a toroidal magnetic field $B_\varphi$ by shearing a pre-existing constant poloidal field 
$\mathbf{B}_{\rm p} = \{B_r,\,B_\theta,\,0\}$ (we use the spherical polar coordinates).
As a result of its buoyancy-related undular instability (e.g., \citealt{a78,svb82}), the toroidal field is prompted
to form magnetic flux rings that will rise toward the bottom of convective envelope, thus producing
chemical mixing and also participating in angular momentum redistribution.

At the stellar equator ($\theta = 90^\circ$), the momentum and induction equations (\citealt{chmg93}) can be reduced to
\bea
\label{eq:mom}
\frac{\partial\Omega}{\partial t} & = & \omega_{{\rm A},r}^2\,\frac{\partial b}{\partial r}, \\
\frac{\partial b}{\partial t} & = & r^2\,\frac{\partial\Omega}{\partial r},
\label{eq:ind}
\eea
where $\omega_{{\rm A},r} = B_r/(\sqrt{4\pi\rho}\,r)$ is the local Alfv\'{e}n frequency associated with the radial field component $B_r$, and $b = rB_\varphi/B_r$.
In equations (\ref{eq:mom}\,--\,\ref{eq:ind}), we have omitted the viscosity and magnetic diffusivity, for these
will be shown to work on much longer time scales than the formation and buoyant rise of magnetic rings.
For the sake of simplicity, we have additionally assumed that the poloidal field's configuration is such that $B_\theta = 0$ at the equator.
Neglecting changes with the radius of the coefficients in equations (\ref{eq:mom}\,--\,\ref{eq:ind}), we find that
the variations of $\Omega$ and $b$ can locally be described by the same Alfv\'{e}n wave equation
$$
\frac{\partial^2 f}{\partial r^2} - \frac{1}{(r^2\omega_{{\rm A},r}^2)}\,\frac{\partial^2 f}{\partial t^2} = 0.
$$
So, at a given radius, the initial (continuing while $t\ll \omega_{{\rm A},r}^{-1}$) decrease of $\Omega$ and increase of $|b|$ 
can be approximated as $\Omega(t,r)\approx\Omega_{\rm max}(r)\cos (\omega_{{\rm A},r}\,t)$, and
$b(t,r)\approx b_{\rm max}(r)\sin (\omega_{{\rm A},r}\,t)$.
From (\ref{eq:mom}\,--\,\ref{eq:ind}) we can also estimate the ratio of the wave amplitudes
$\Omega_{\rm max}/b_{\rm max}\approx \omega_{{\rm A},r}/r$, hence $(B_\varphi)_{\rm max}\approx B_r(\Omega_{\rm max}/\omega_{{\rm A},r})$.
This ensures that the sum of the rotational kinetic and toroidal field potential energy (per unit gram) is conserved,
\bea
\label{eq:energy}
\frac{1}{2}\left(r\Omega\right)^2 + \frac{B_\varphi^2}{8\pi\rho} = \frac{1}{2}\left(r\Omega_{\rm max}\right)^2.
\eea

Substituting the values of $r=r_{\rm mix}$ and $\rho=\rho\,(r_{\rm mix})$ from our RGB model (Table~\ref{tab:tab1}) into
the above relations, we find that
\bea
\label{eq:tbphi}
\omega_{{\rm A},r}^{-1} & = & 1.50\times 10^3\,B_r^{-1}\ \ \mbox{yr\,rad}^{-1}, \\
\label{eq:bphi0}
(B_\varphi)_{\rm max} & = & 4.75\times 10^5\,(\Omega_{\rm max})_{-5}\,B_r,
\eea
where $(\Omega_{\rm max})_{-5} \equiv \Omega_{\rm max}/(10^{-5}\,\mbox{rad}\,\mbox{s}^{-1})$.
Equation (\ref{eq:bphi0}) gives an order-of-magnitude estimate of the toroidal field maximum 
strength that may be generated by the differential rotation at $r=r_{\rm mix}$, while equation (\ref{eq:tbphi}) estimates
its growth time. For example, we may expect (from equation \ref{eq:ind}) that, after a time $\Delta t\ll \omega_{{\rm A},r}^{-1}$,
a toroidal field $B_\varphi\approx \Omega_{\rm max}B_r\,q\,\Delta t \approx (B_\varphi)_{\rm max}\, \omega_{{\rm A},r}\,q\,\Delta t 
= 3.16\times 10^2(\Omega_{\rm max})_{-5} B_r\,q\,\Delta t$
will be created, where $q = (\partial\ln\Omega/\partial\ln r) < 0$ is the initial rotational shear, and $\Delta t$ is expressed in years. Assuming that 
$\Omega_{\rm max} = \Omega(r_{\rm mix}) = 10^{-3}$\,rad\,s$^{-1}$ (Table~\ref{tab:tab1}) 
at the moment when the differential rotation begins to stretch the poloidal field, we obtain the estimate 
\bea
B_\varphi(\Delta t,\,r_{\rm mix}) \approx 3.16\times 10^4 B_r\,q\,\Delta t,
\label{eq:bphiin1yr}
\eea
where $\Delta t \ll 1.50\times 10^3\,B_r^{-1}$ yrs.
The relative decrease of $\Omega$ for the same period of time is $|\Delta\Omega/\Omega_{\rm max}|\approx 4.42\times 10^{-7}\,B_r^2\,q^2\,(\Delta t)^2\ll 1$.
In the general case of $B_\theta\neq 0$ and $\Omega = \Omega(r,\theta)$ we would have $B_\varphi \approx r\sin\theta\,(\nabla\Omega,\mathbf{B}_{\rm p})\,\Delta t$
for $\Delta t\ll \omega_{{\rm A},r}^{-1}$ (\citealt{s99}), i.e. the both poloidal field components would be involved into the winding up of
toroidal field.

If the uniform rotation of the solar radiative core and the core-envelope rotational coupling in the low-mass MS stars 
are both produced by the back reaction of the azimuthal component of the Lorentz force emerging in the process of
generation of toroidal magnetic fields by the shearing of pre-existing poloidal fields, as proposed by
\cite{chmg93}, then the MS progenitors of low-mass RGB stars are required to possess poloidal magnetic fields with
strengths of the order of 0.01\,G to 10\,G in their radiative interiors. The low magnetic diffusivity $\eta\sim 10^2$\,sm$^2$\,s$^{-1}$
in the low-mass MS stars excludes the ohmic dissipation of these fields. Therefore, the post-MS contraction of the H-exhausted core
from its MS radius of $\sim$\,$0.2\,R_\odot$ down to its RGB radius $r_{\rm c} \approx 0.02\,R_\odot$ and magnetic flux conservation
might lead to the amplification of $B_r$ up to the values 1\,G to 1\,kG. From the same considerations,
it also follows that $B_r$ in the radiative zone of the RGB star may decrease with the radius as $B_r\approx B_r(r_{\rm mix})\,(r_{\rm mix}/r)^2$.
It is true that the mass inflow in the radiative zone may sweep the frozen-in poloidal field toward the H burning shell
on a long time scale of the order of $\Delta r/|\dot{r}| = (r_{\rm bce}-r_{\rm mix})/|\dot{r}|\sim 3\times 10^7$ yrs (Table~\ref{tab:tab1}). 
However, the poloidal field
may be replenished by a dynamo operating at the bottom of convective envelope. It may then be entrained and redistributed all over
the radiative zone by the mass inflow, which could also produce the dependence $B_r\propto r^{-2}$.
These estimates will be used in our further analysis.

\section{The Buoyant Rise of Magnetic Flux Rings}
\label{sec:ringmotion}

\subsection{General Results}

Because we assume a differential rotation in the radiative zone, the inertial frame of reference has been chosen.
The magnetic flux rings are assumed to be axisymmetric with respect to the rotation axis.
In the spherical polar coordinates their radial and latitudinal accelerations are
\bea
\label{eq:durdt}
\frac{du_r}{dt} & = & \frac{u_{\theta}^2}{r}+\left(\frac{u_{\varphi}^2}{r}-\Omega^2r\sin^2\theta\right)+
\frac{\rho_{\rm e}-\rho}{\rho_{\rm e}+\rho}\left(\frac{GM_r}{r^2}-\Omega^2r\sin^2\theta\right) \nonumber \\
& & -\frac{B_{\varphi}^2}{4\pi r(\rho_{\rm e}+\rho)} - \frac{C_{\rm D}}{\pi a}\frac{\rho_{\rm e}}{\rho_{\rm e}+\rho}u_r\sqrt{u_r^2+u_{\theta}^2}, \\
\label{eq:duthetadt}
\frac{du_\theta}{dt} & = & -\frac{u_ru_\theta}{r}+\left(\frac{u_{\varphi}^2}{r}\cot\theta - \Omega^2r\sin\theta\cos\theta\right)
- \frac{\rho_{\rm e}-\rho}{\rho_{\rm e}+\rho}\,\,\Omega^2r\sin\theta\cos\theta \nonumber \\
& & - \frac{B_{\varphi}^2}{4\pi r(\rho_{\rm e}+\rho)}\cot\theta - \frac{C_{\rm D}}{\pi a}\frac{\rho_{\rm e}}{\rho_{\rm e}+\rho}u_\theta\sqrt{u_r^2+u_{\theta}^2}.
\eea
These equations (\citealt{chg87,mgc03,mdm04}) take into account the centrifugal reduction of the local gravitational acceleration, 
the buoyant force, the magnetic tension force, and the aerodynamic drag force (we employ the drag coefficient $C_{\rm D} = 1$). 
They are supplemented with the equation
\bea
\label{eq:duphidt}
\frac{d}{dt}(u_\varphi r\sin\theta) = 0
\eea
describing the conservation of the azimuthal component of the specific angular momentum. The subscript ``e'' means that the respective quantity
is referred to the external medium surrounding the rings, while all other quantities, except $\Omega$, refer to ring properties.

At its starting position $(r_0,\,\theta_0)$, a ring is specified by its initial cross-section radius $a_0$, the strength of the frozen-in
toroidal magnetic field $B_{\varphi,0}$, and internal thermodynamic properties: $P_0$, $T_0$, $\rho_0$, and $\mu_0$. The dynamic equilibrium
between the ring and its surrounding medium requires that
\bea
P_{\rm e} = P + \frac{B_{\varphi}^2}{8\pi},\ \ \mbox{or}\ \ \frac{\Delta P}{P_{\rm e}} \equiv \frac{(P_{\rm e}-P)}{P_{\rm e}} = 
\frac{1}{(1+\beta)}\approx\frac{1}{\beta},
\label{eq:p}
\eea
where $\beta\equiv P/(B_\varphi^2/8\pi)\gg 1$ is the ratio of the thermodynamic to magnetic pressure in the ring.
The conservation of mass and magnetic flux of the ring determine how its radius and toroidal field evolve during its motion
\bea
\left(\frac{a}{a_0}\right)^2 & = & \frac{\rho_0r_0\sin\theta_0}{\rho r\sin\theta} = \frac{B_{\varphi,0}}{B_\varphi},
\label{eq:conserv}
\eea
where $r_0 \leq r\leq r_{\rm bce}$.
For the equation of state, we use the ideal gas law. Finally, the entropy change in the ring is described by the simple equation
\bea
\frac{dS}{dt} = \frac{32(\gamma-1)\sigma T_{\rm e}^4}{3\kappa_{\rm e}\rho_{\rm e}a^2P}\,\delta T =
2\gamma K\,\frac{\delta T}{a^2}
\label{eq:dsdt}
\eea
derived by \cite{mgc03}. Here, $S=\ln\,[(P/P_0)(\rho_0/\rho)^\gamma]$,  $\delta T = (T_{\rm e} - T)/T_{\rm e}$, and $\gamma = 5/3$.

The initial conditions for the ring are specified by $r_0=r_{\rm mix} = 0.05\,R_\odot$, $0 < \theta_0\leq 90^\circ$,
$u_r = 0$, $u_\theta = 0$, $u_\varphi = r_0\sin\theta_0\,\Omega(r_0,\,\theta_0)$, $\beta =\beta_0$,
$a = a_0$, $S=0$, $\delta P\equiv (P_{\rm e} - P)/P_{\rm e} = (1+\beta_0)^{-1}$,
$\delta\rho\equiv (\rho_{\rm e} - \rho)/\rho_{\rm e} = (1+\beta_0)^{-1}$, and $\delta T=0$.
The ring's radius $a$ will be measured in units of the local pressure scale height $H_P = P_{\rm e}/g\rho_{\rm e}$, where
$g=GM_r/r^2$ is the gravitational acceleration. 
The implicit assumption of uniform pressure inside the ring requires that $a\ll H_P$ (the thin ring approximation).
For brevity, we will denote $a\equiv (a/H_P)$ and, given the previous remark, we will only consider cases with $a\ll 1$. 

\begin{figure}
\plotone{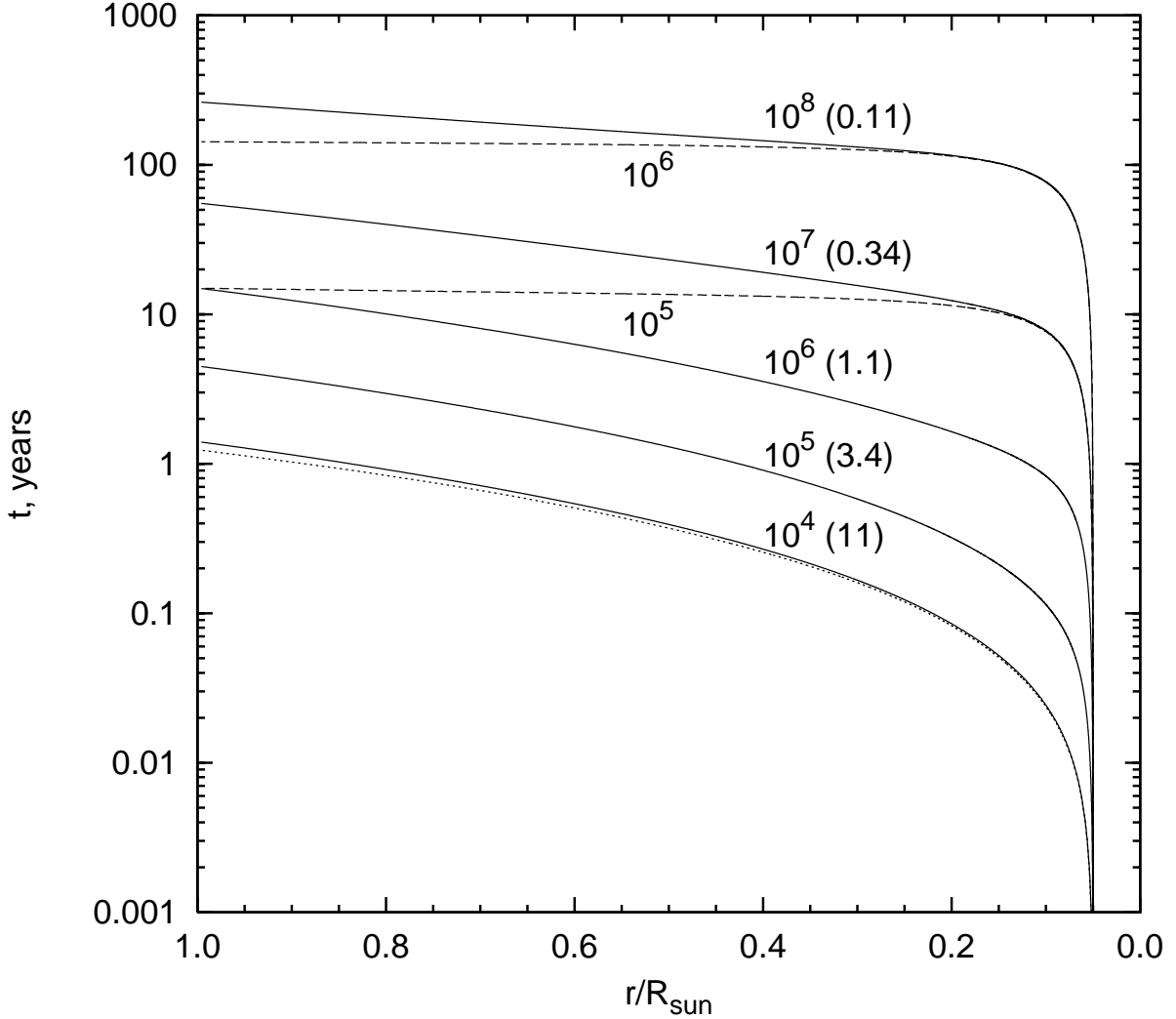}
\caption{The motion of the equatorial magnetic flux rings with the cross-section radii $a_0 = 10^{-4}\,H_P$
         (solid and dotted curves) and $a_0 = 10^{-3}\,H_P$ (dashed curves). The powers of ten show
         the specified values of $\beta_0$, while the numbers in parentheses give the corresponding
         strengths (in MG) of the frozen-in toroidal magnetic field. The dotted curve corresponds
         to the ring with $\beta_0 = 10^{10}$ ($B_\varphi\approx 11$ kG) whose mean molecular weight
         has been reduced by $\Delta\mu = 5\times 10^{-5}$.
         }
\label{fig:f2}
\end{figure}

The above differential and algebraic equations have been solved numerically. The motion of the ring in the equatorial plane ($\theta_0 = 90^\circ$)
is shown in Fig.~\ref{fig:f2} for our assumed rotation law $\Omega(r,\theta) = \Omega(r_{\rm mix})(r_{\rm mix}/r)^2$, in which case $\theta(t)=\theta_0$. Solid curves
correspond to the initial value of $a_0 = 10^{-4}$, while dashed curves have $a_0 = 10^{-3}$. The powers of ten
near the curves give their specified values of $\beta_0$, the respective toroidal field strengths (in MG) being displayed in parentheses.
Our computations have shown that for $\beta_0\ga 10^8$ (in other words, for  $(B_\varphi)_0 \la 0.1$ MG) 
and $a_0 \ga 2\times 10^{-4}$ the dependence of the total buoyant rising time on $\beta_0$ and $a_0$ can be approximated as
\bea
t_{\rm b} \approx 1.4\times 10^2\,\left(\frac{\beta_0}{10^8}\right)\, \left(\frac{a_0}{10^{-4}}\right)^2\ \ \mbox{yr}.
\label{eq:tb}
\eea
This is transformed into the average buoyant velocity
\bea
\langle v_{\rm b} \rangle = \frac{\Delta r}{t_{\rm b}} = \frac{(r_{\rm bce} - r_{\rm mix})}{t_{\rm b}} \approx
15\,\left(\frac{10^8}{\beta_0}\right)\,\left(\frac{10^{-4}}{a_0}\right)^2\ \ \mbox{cm\,s}^{-1},
\label{eq:vb}
\eea
which is nearly five orders of magnitude smaller than the appropriately scaled average velocity used by BWNC
(for the scaling, we have used the data from the fifth column of Table~\ref{tab:tab1} and the column RGB-1 of Table~\ref{tab:tab2}).
We have tested that this big difference is entirely caused by the assumption of thermal equilibrium between the ring and its surroundings
made in the cited paper.

\begin{figure}
\plotone{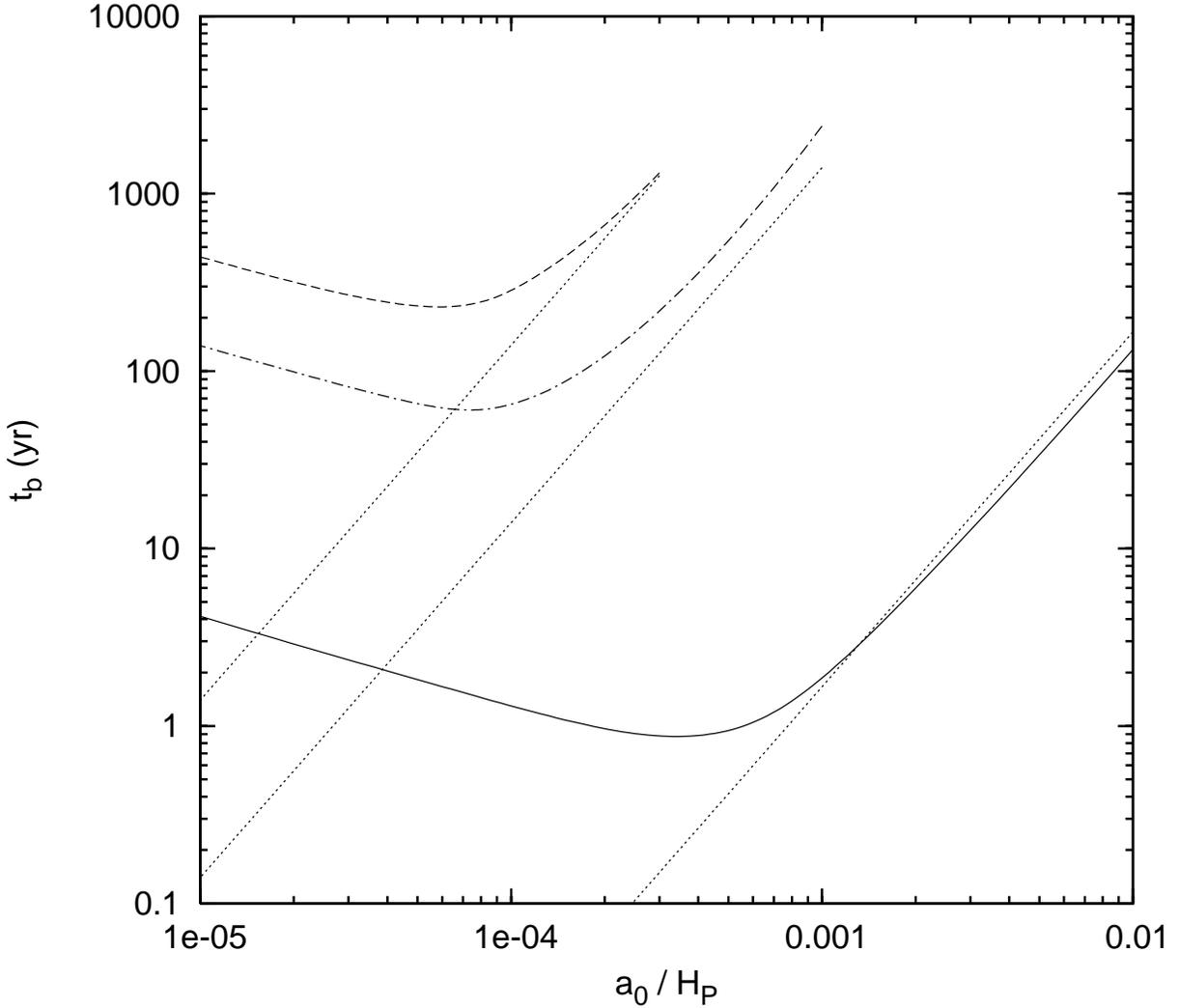}
\caption{The dashed and dot-dashed curves show the dependences of the buoyant rising time on the ring
         cross-section radius for the rings with $\beta_0 = 10^8$ and $\beta_0 = 10^7$, respectively,
         whose chemical composition has not been changed by nuclear reactions ($\Delta\mu = \mu_{\rm e} - \mu = 0$),
         while the solid curve corresponds to the ring with $\beta_0 = 10^{10}$ and $\Delta\mu = 5\times 10^{-5}$
         ($\mu$ reduced by $^3$He burning).
         These three curves represent our numerical solutions of equations (\ref{eq:durdt}\,--\,\ref{eq:duphidt}).
         The dotted lines give their corresponding approximations by equation (\ref{eq:tb}), in which
         $\beta_0$ has been replaced by $\beta_{\rm eff} = \delta\mu^{-1} = (\mu/\Delta\mu)$ for the third case.
         }
\label{fig:f3}
\end{figure}

The dashed and dot-dashed curves in Fig.~\ref{fig:f3} show the dependences of $t_{\rm b}$ on $a_0$ for $\beta_0 = 10^8$ 
and $\beta_0 = 10^7$ obtained by solving equations (\ref{eq:durdt}\,--\,\ref{eq:duphidt}), while the dotted lines
represent their approximations by (\ref{eq:tb}). It is seen that the exact solutions strongly deviate from the approximate ones
at $a_0 \la 2\times 10^{-4}$. This limit corresponds to a regime in which the aerodynamic drag force comes into play.
Indeed, the last terms on the right-hand-sides of equations (\ref{eq:durdt}\,--\,\ref{eq:duthetadt}), that describe the drag
force, are negative and inversely proportional to $a$. They lead to a slowing down of the ring's buoyant rise at small $a$,
which is not reflected in equation (\ref{eq:tb}).

\subsection{Rings with $\mu$ Reduced by $^3$He Burning}
\label{sec:mured}

Our numerical computations can optionally take into account the fact that the rings carrying the nuclear processed material
are necessarily formed in the region of the local $\mu$ depression maintained by the $^3$He burning (Fig.~\ref{fig:f1}b).
They must therefore have a lower $\mu$ than the bulk of the radiative zone through which they move. 
As a result, their average buoyant velocity is found to weakly depend on the toroidal field
strength. However, the frozen-in toroidal field is still needed for them to remain cohesive
while rising.

Dotted curve in Fig.~\ref{fig:f2} shows the path of a ring with 
$\Delta\mu = \mu_{\rm e} - \mu = 5\times 10^{-5}$,
$a_0 = 10^{-4}$, and $\beta_0 = 10^{10}$ that corresponds to $(B_\varphi)_0 \approx 11$ kG. Its buoyant rising time is
$t_{\rm b}\approx 1.3$ yrs. Approximately the same short time is obtained for a ring with
$(B_\varphi)_0 \approx 1.1$ kG ($\beta_0 = 10^{12}$). If those two rings had $\mu = \mu_{\rm e}$ then their rising times would be much longer
and vastly different, namely: $t_{\rm b}\approx 10^4$ yrs, and $t_{\rm b}\approx 10^6$ yrs, respectively (eq. \ref{eq:tb}).
This means that their buoyancy is now controlled by the $\mu$ difference alone.
Let us neglect, for a moment, the difference in temperature between the surrounding medium and the ring,
$\delta T\equiv (T_{\rm e}-T)/T_{\rm e} \approx 0$. The buoyant acceleration is
$a_{\rm b} \approx g\delta\rho$. In the case of $\Delta\mu = 0$, the initial positive difference in density 
$\delta\rho \approx\delta P\approx \beta_0^{-1}$
is simply due to the excess magnetic pressure inside the ring (eq. \ref{eq:p}). On the other hand, in the case of
$\Delta\mu > 0$, and for a weak toroidal field ($\delta P\approx 0$), 
we have $\delta\rho\approx\delta\mu\equiv\Delta\mu/\mu_{\rm e}$. We can introduce
the effective toroidal magnetic field $B_{\rm eff}$ associated with a specified value of $\delta\mu$ such that
$\delta\mu = \beta_{\rm eff}^{-1}$. It turns out that $B_{\rm eff}\approx 9.8$ MG for $\Delta\mu = 5\times 10^{-5}$ at
$r=r_{\rm mix}$ in our RGB model. Hence, toroidal magnetic fields with $(B_\varphi)_0\ll B_{\rm eff}$ will have a negligible effect
on the motion of the ring with the reduced $\mu$. For $\Delta\mu = 5\times 10^{-5}$, $\beta_0 = 10^{10}$, and $a_0 = 10^{-3}$
the rising time $t_{\rm b}\approx 1.6$ yrs is still short. However, it increases up to 
$t_{\rm b}\approx 33$ yrs and $t_{\rm b}\approx 132$ yrs for rings with the initial radii $a_0 = 5\times 10^{-3}$ and $a_0 = 10^{-2}$, respectively,
following the dependence $t_{\rm b}\propto a_0^2$ for the thicker rings (eq. \ref{eq:tb}).

The role that the reduced $\mu$ plays in the acceleration of the buoyant rise of the ring containing a weak magnetic field can be elucidated
if we consider, for simplicity, that the ring's vertical motion consists of the following recurring sequence. Initially, let the ring
have $\delta T=0$ but $\delta\mu > 0$, hence $\delta\rho = \delta\mu + \beta_0^{-1} >0$. Now, let the ring rise adiabatically
until the accumulated difference in $T$ compensates that in $\rho$, i.e. $\delta T = \delta\rho$. At this moment, the ring stops and waits a while   
for the heat exchange to make $\delta T = 0$ before starting to rise adiabatically again, and so on. But the waiting time is in fact 
the thermal time $\tau_{\rm th}$, therefore it is inversely proportional to
the heating rate $dS/dt$ that linearly depends on the ratio $\delta T/a^2$ (eq. \ref{eq:dsdt}). When the ring stops, it has $\delta T = \delta\mu + \beta_0^{-1}$.
Therefore, if $\delta\mu \gg \beta_0^{-1}$ then $\delta T\approx \delta\mu$, hence we have 
$t_{\rm b}\propto\tau_{\rm th}\propto\delta\mu^{-1}\,a^2 = \beta_{\rm eff}\,a^2$. On the other hand,
if $\delta\mu \ll \beta_0^{-1}$ then $\delta T \approx \beta_0^{-1}$, and $t_{\rm b}\propto\tau_{\rm th}\propto \beta_0\,a^2$
(eq. \ref{eq:tb}). The transition from the one to the other regime occurs at $\beta_0^{-1}\approx\delta\mu = \beta_{\rm eff}^{-1}$.

We have checked that equation (\ref{eq:tb}) gives a correct order-of-magnitude estimate 
(this time, at $a_0\ga 8\times 10^{-4}$ though) for the buoyant rising time of
a ring with a reduced $\mu$ provided that $\beta_0$ is replaced by $\beta_{\rm eff} = \delta\mu^{-1}$ (the dotted line approximating
the solid curve in Fig.~\ref{fig:f3}). It is also important to note
that the coefficients in equations (\ref{eq:tb}\,--\,\ref{eq:vb}) have been obtained for a particular RGB model (our model
from Table~\ref{tab:tab1}), therefore they are model dependent. For example, given that $t_{\rm b}$ is expected to be
inversely proportional to the thermal diffusivity $K$ as well (eq. \ref{eq:dsdt}), the latter being roughly proportional to
the luminosity, we predict that these coefficients should change by a factor of 10 ($t_{\rm b}$ decreases, while $\langle v_{\rm b}\rangle$
increases) toward the RGB tip ($\log\,L/L_\odot\approx 3.3$). This scaling is indeed confirmed by our computations.

\subsection{Comparison with Results Obtained by BWNC}

Given that one rising magnetic flux ring carries the mass $m_{\rm b}=2\pi^2r_0\,a_0^2\,\rho_0\sin\theta_0=2\pi^2ra^2\rho\sin\theta$, $N$ such rings
present in the radiative zone at the same time will provide chemical mixing with the mass rate
\bea
\dot{M}_{\rm b} = \frac{Nm_{\rm b}}{t_{\rm b}},
\label{eq:dmdt}
\eea
where $t_{\rm b}$ is their buoyant rising time. The quantity $\dot{M}_{\rm b}$ has to match the observationally constrained
rate of the RGB extra mixing $\dot{M}_{\rm mix}\approx 4\times 10^{-8}\,M_\odot$\,yr$^{-1}$ (BWNC).
The right-hand-side of equation (\ref{eq:dmdt}) is a function of the ring parameters $N$, $a_0$, and $(B_\varphi)_0$.
For a ring with a reduced $\mu$, the third parameter should be replaced with $\Delta\mu$, unless $\delta\mu\ll 1/\beta_0$ (\sect{sec:mured}).
In the case of $\Delta\mu = 0$, considered by BWNC, the equating of $\dot{M}_{\rm b}$ to $\dot{M}_{\rm mix}$
gives a relationship between $N$, $a_0$, and $(B_\varphi)_0$. To estimate a reasonable value of $a_0$ at $r=r_{\rm mix}$, they have referred to
the characteristic dimension $a_\odot\sim$\,1000\,--\,2000 km of magnetic flux tubes that are believed to exist deep in the solar convective zone.
Beside the RGB model, they have also considered a representative model for low-mass asymptotic giant branch (AGB) stars, whose structure above the H burning shell
resembles that of upper RGB stars. There is indirect evidence, such as distinctive $^{18}$O/$^{16}$O, $^{17}$O/$^{16}$O, $^{12}$C/$^{13}$C, and
N/C abundance ratios in the meteorite grains of AGB circumstellar origin (\citealt{wea06}) and in the atmospheres of carbon-enhanced metal-poor stars 
(\citealt{rea05,sea06,mea06,dp08a,lea08}), indicating that extra mixing may operate in the radiative zones of these stars too (\citealt{nbw03}). In their RGB-1
and RGB-2 magnetic buoyancy models, BWNC have employed the values of $a_0$ obtained 
assuming that $a(r_{\rm bce})=a_\odot$ and using equation (\ref{eq:conserv}) with the stellar structure parameters
from their AGB and RGB stellar models, respectively. These values are listed in our Table~\ref{tab:tab2}
along with the corresponding estimates for our RGB model.

BWNC have also assumed that there is only one magnetic flux ring floating in the radiative zone at any time. 
Substituting the number $N=1$ together with the values of $a_0$ from
Table~\ref{tab:tab2} and $\dot{M}_{\rm mix}=4\times 10^{-8}\,M_\odot$\,yr$^{-1}$ into equation (\ref{eq:dmdt}) constrains
the required values for $t_{\rm b}$ and $\langle v_{\rm b}\rangle$ (Table~\ref{tab:tab2}). 
Under the assumptions, made by BWNC, that the ring always stays in the thermal equilibrium with
its surrounding medium ($\delta T =0$) and that the only force impeding its motion is the aerodynamic drag force, it is easy to show that
$t_{\rm b}\propto \sqrt{\beta_0/a_0}\propto (B_\varphi)_0/\sqrt{a_0}$. 
It is from the last relation that BWNC have determined the toroidal field strengths 
needed to drive the RGB extra mixing by magnetic buoyancy (Table~\ref{tab:tab2}, $(B_\varphi)_0$ is transformed 
to $(B_\varphi)_{\rm bce}$ using equation \ref{eq:conserv}).
They pointed out that these results are consistent with existing observations of magnetic fields in red giants (e.g., \citealt{bea01}).

However, as we have noted, BWNC underestimated the ring's rising time by assuming that the heat exchange
between the surrounding medium and the ring occurs instantaneously, which maintains $\Delta T = 0$ all the time.
Our more conservative assumption explicitly takes into consideration the radiative heat exchange, which
leads to a much longer ring's rising time, as approximated by equation (\ref{eq:tb}). Employing the values of
$a_0$, $t_{\rm b}$, and $\beta_0$ from the columns RGB-1 and RGB-2 of Table~\ref{tab:tab2}, we find that the rising times have
been underestimated by the factors $6\times 10^4$ and $4\times 10^6$ for these buoyancy models. To keep
the same estimates for the toroidal field strength, the ring number $N$ has to be increased by the corresponding factors.
However, the volume occupied by $\sim$\,$10^5$\,--\,$10^6$ rings is comparable to the total volume of
the radiative zone, in which case the stellar structure would be greatly disturbed, especially at the bottom of
convective envelope where the rings encounter turbulent convection. Here, the frozen-in toroidal field
is either quickly dissipated via the strong turbulent diffusion, thus depositing its energy at the radiative/convective
interface, or it inhibits convective motions if its potential energy exceeds the turbulent kinetic energy 
(\citealt{moss03}). In either case, the stellar structure would be strongly modified.
In the next section, we will show that a reasonably small number of $N$ can be obtained only for
rings with a reduced $\mu$.

\subsection{There Still May Be a Solution}

\begin{figure}
\epsfxsize=12cm
\epsffile [30 130 400 700] {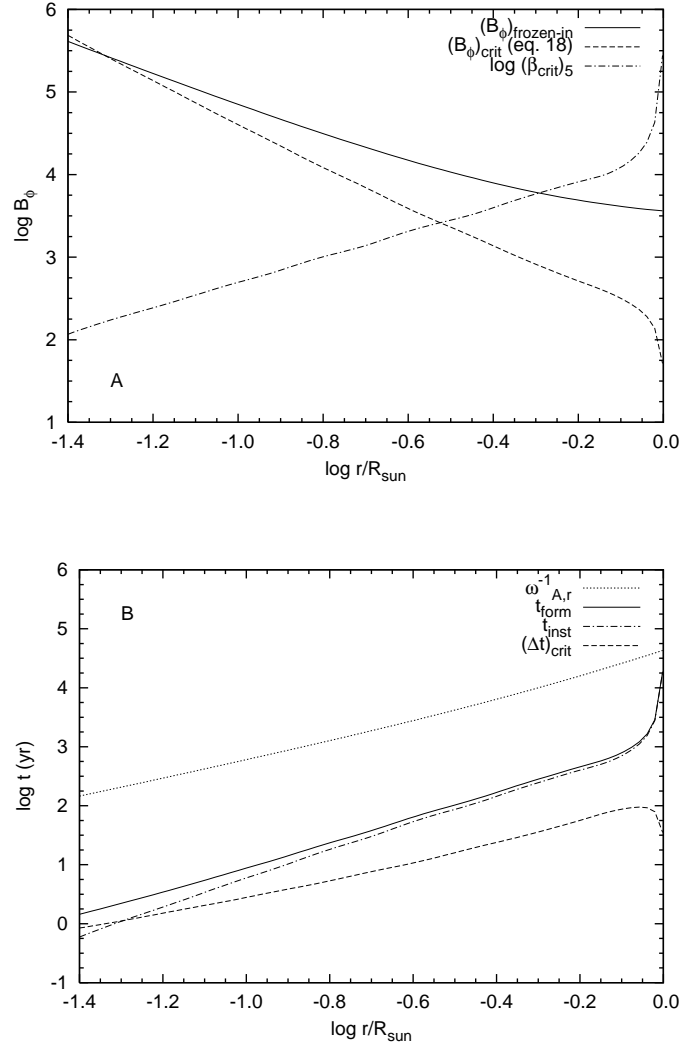}
\caption{(a) The critical toroidal field (dashed curve) and parameter $\beta$ (dot-dashed curve) for the triggering of the buoyancy instability
         (equation \ref{eq:bcrit}), and the evolution of the toroidal field strength in the rising ring
         (solid curve, equation \ref{eq:conserv}). (b) The time scales for the winding up of the critical
         toroidal filed ($\Delta t_{\rm crit}$) and for the development of the buoyancy instability ($t_{\rm inst}$),
         as well as the total ring formation time ($t_{\rm form} = \Delta t_{\rm crit} + t_{\rm inst}$).
         These time scales should be shorter than the toroidal field growth time $\omega_{{\rm A},r}^{-1}$.
         In these computations, it has been assumed that $B_r = 7.32\,(r_{\rm mix}/r)^2$ G, which makes
         $(\Delta t)_{\rm crit} = t_{\rm inst}$ at $r = r_{\rm mix}$, where $\log (r_{\rm mix}/R_\odot)\approx -1.30$.
         }
\label{fig:f4}
\end{figure}

By assigning $N=1$, BWNC have implicitly assumed that the average time $t_{\rm form}$ needed to form
a magnetic flux ring is equal to its buoyant rising time $t_{\rm b}$. Indeed, instead of  (\ref{eq:dmdt}) the buoyancy mixing mass rate
ought to be calculated as $\dot{M}_{\rm b} = m_{\rm b}/t_{\rm form} = Nm_{\rm b}/t_{\rm b}$, where $N = t_{\rm b}/t_{\rm form}$.
We assume that, like in the case of the solar tachocline (e.g., \citealt{sr83}), 
the appropriate MHD mechanism responsible for the formation of magnetic flux rings in the vicinity of the H burning shell is the undular buoyancy instability.
The criterion for its development has been extensively discussed in the literature, e.g. by \cite{a78}, \cite{svb82}, \cite{s99}, and \cite{f01}.
It has been shown that a diffusive toroidal magnetic field gets broken into distinct arching flux tubes when
\bea
B_\varphi > (B_\varphi)_{\rm crit} \approx \sqrt{4\pi\rho\,r^2 N^2\,\frac{H_P}{r}\,\frac{\eta}{K}},
\label{eq:bcrit}
\eea
provided that $(\partial\ln B_\varphi/\partial\ln r) = O(1)$. A profile of $(B_\varphi)_{\rm crit}$ in the radiative zone of our RGB model is
plotted with the dashed curve in Fig.~\ref{fig:f4}a. Its corresponding profile of $\beta_{\rm crit}$ is shown with
the dot-dashed curve. At $r=r_{\rm mix}$ ($\log\,r_{\rm mix}/R_\odot\approx -1.30$), the critical toroidal field is 254 kG
and $\beta_{\rm crit} = 1.8\times 10^7$ (Table~\ref{tab:tab2}).

In the presence of rapid rotation, such that $\Omega\gg\omega_{{\rm A},\varphi} \equiv B_\varphi/(\sqrt{4\pi\rho}\,r)$,
the Coriolis force reduces the instability's growth rate $\omega_{{\rm A},\varphi}$ by the factor $\omega_{{\rm A},\varphi}/\Omega$ (\citealt{pt85,s99}).
In our RGB model, $\Omega(r_{\rm mix}) = 10^{-3}$\,rad\,s$^{-1}$ strongly exceeds $\omega_{{\rm A},\varphi}\,(r_{\rm mix}) = 2.11\times 10^{-11}\,B_\varphi$\,rad\,s$^{-1}$
for all reasonable strengths of $B_\varphi\ll 47$ MG. Therefore, we use
the estimate of $t_{\rm inst}\approx \Omega/\omega_{{\rm A},\varphi}^2$ for the instability growth time. 
The winding up of a toroidal field by differential rotation continues until $B_\varphi\approx \Omega B_r\Delta t\,q$ 
reaches the critical value (\ref{eq:bcrit}). 
This takes $(\Delta t)_{\rm crit}\approx 3.17\times 10^{-3}\,(B_\varphi)_{\rm crit}/(B_r\,\Omega_{-5}\,q)$ years.
After that, it will take another $t_{\rm inst}\approx 1.93\times 10^{10}\,\left[\,\rho\,(r/R_\odot)^2\,\Omega_{-5}\right]/(B_\varphi)_{\rm crit}^2$ years
for the buoyancy instability to occur and assemble the rings. So, the total ring formation time is $t_{\rm form} = (\Delta t)_{\rm crit} + t_{\rm inst}$.
For our simple estimates to be true, $t_{\rm form}$ should be shorter than 
$\omega_{{\rm A},r}^{-1} = 7.82\times 10^3\,\sqrt{\rho}\,(r/R_\odot)/B_r$ yrs, which allows to consider that the rotational shear
is nearly constant and that, for our assumed rotation law in the radiative zone $\Omega_{-5} = 10^2\,(r_{\rm mix}/r)^2$, 
$q = O(1)$. For this case, and assuming that
$B_r = 7.32\,(r_{\rm mix}/r)^2$ G, the four characteristic time scales are plotted in Fig.~\ref{fig:f4}b.
In particular, we have $t_{\rm form} = 2.20$ yrs at $r = r_{\rm mix}$.
Note that $(\Delta t)_{\rm crit}\propto (B_r\Omega)^{-1}$, $t_{\rm inst}\propto\Omega$, and
$\omega_{{\rm A},r}^{-1}\propto B_r^{-1}$.
We have used the parameter $B_r(r_{\rm mix}) = 7.32$ G for which
the time scales $(\Delta t)_{\rm crit}$ and $t_{\rm inst}$ coincide.
Its value scales as $\Omega^{-2}(r_{\rm mix})$. At a fixed value of $\Omega(r_{\rm mix})$, $B_r(r_{\rm mix})$
determines which of the two time scales, $(\Delta t)_{\rm crit}$ or $t_{\rm inst}$, makes a predominant
contribution to $t_{\rm form}$.

It is very likely that the buoyancy instability forms not just one but a number $n > 1$ of magnetic flux rings during the time $t_{\rm form}$,
therefore $N = n(t_{\rm b}/t_{\rm form})$.
The equating of $\dot{M}_{\rm b}=n m_{\rm b}/t_{\rm form}$ to $\dot{M}_{\rm mix} = 4\times 10^{-8}\,M_\odot$\,yr$^{-1}$ constrains
the ring's minimum radius as a function of $t_{\rm form}$, $n$, and the initial colatitude 
\bea
\left(\frac{a_0}{H_P}\right)^2 = \frac{\dot{M}_{\rm mix}\,t_{\rm form}}{n\,2\pi^2r_0^3\rho_0\sin\theta_0}\left(\frac{H_P}{r_0}\right)^{-2} =
9.78\times 10^{-5}\,\frac{t_{\rm form}}{n\,\sin\theta_0},
\label{eq:a0}
\eea
where $r_0 = r_{\rm mix}$, and $t_{\rm form}$ is expressed in years.

For the values of $t_{\rm form} = 2.20$ yrs, $n = 1$, and $\theta_0 = 90^\circ$, we calculate $a_0=1.47\times 10^{-2}$ (in units of $H_P$, as usually), and the total
number of rings in the radiative zone $N = 0.454\,t_{\rm b}$. 
As we have mentioned before, the ring with the radius $a_0 = 10^{-2}$ and $\mu$ reduced by $\Delta\mu = 5\times 10^{-5}$ has the buoyant rising time $t_{\rm b}\approx 132$ yrs.
This means that the ring with $a_0 = 1.47\times 10^{-2}$ and the same mean molecular weight would cross the radiative zone
in $t_{\rm b}\approx (1.47)^2\times 132 = 285$ yrs (eq. \ref{eq:tb}). Hence, we have $N = 129$.
Note that this number does not change if the values of the parameters $t_{\rm form}$ and $n$ are varied, provided that
our buoyancy mixing mass rate is still constrained to match $\dot{M}_{\rm mix}$. Indeed, in this case
both the ratio $t_{\rm form}/n$ (eq. \ref{eq:a0}) and $t_{\rm b}$ (eq. \ref{eq:tb}) 
are proportional to $a_0^2$ which leads to its cancellation in the relationship $N = n(t_{\rm b}/t_{\rm form})$.

We have shown that the reduced mean molecular weight in the magnetic flux rings formed in the region of $^3$He burning
accelerates their buoyant rise quickly enough for their total number needed to maintain the RGB extra mixing to be reasonably small
($N\approx 10^2$ rings would only occupy $\sim$\,$10^{-3}$\,--\,$10^{-4}$ part of the radiative zone).
However, the estimated value of $N\approx 1.3\times 10^2$ is in fact a lower limit obtained under the most favourable
assumptions. In particular, we have silently assumed that all rings have the same formation time and cross-section radius related by
equation (\ref{eq:a0}), and that no rings are formed above the radius $r_{\rm mix}$. These assumptions are obviously not true.
Relaxing either of them will certainly increase the total ring number in the radiative zone.
To figure out what effect may be produced by relaxing the first assumption we would have to determine 
a relationship between $t_{\rm form}$ and $a_0$ for a spectrum of rings created by the buoyancy instability, e.g. like it has been done
for the solar tachocline by \cite{sr83}. This problem is out of scope of the present preliminary study.

We find it more important to address here the second issue. Indeed, given that the critical strength of
toroidal magnetic field for triggering the buoyancy instability decreases rapidly with the radius
(dashed curve in Fig.~\ref{fig:f4}a), while our estimated ring formation time stays much shorter than
the toroidal field growth time $\omega^{-1}_{{\rm A},r}$ (Fig.~\ref{fig:f4}b),
the formation of magnetic flux rings at $r_0 > r_{\rm mix}$ appears to be unavoidable. These rings
would contribute to the total ring number present in the radiative zone at the same time but they would not
participate in the chemical element transport because their constituent material has not been nuclearly processed.
In fact, the buoyancy of these "parasitic" rings should be reduced compared to the buoyancy of the rings
originating at $r\approx r_{\rm mix}$ because their mean molecular weight does not differ
from that of the surrounding medium through which they rise. On the other hand, they are formed
in a region where the thermal diffusivity $K$ is higher than at $r = r_{\rm mix}$, therefore their heat exchange
with the surroundings goes faster, which should accelerate their buoyant rise. Our computations show that the rings
formed in the region of constant $\mu$ (at $r\ga r_{\rm mix} + 0.05\,R_\odot$) rise $10^3$ to $10^4$ times
slower than the rings formed in the $\mu$-depression domain. Given that their formation takes between
10 and 100 yrs (solid curve in Fig.~\ref{fig:f4}b), a number of the "parasitic" rings present in the radiative zone
at the same time may be as large as $10^5$.
Unfortunately, our model is too simple to be able to predict if and how the "parasitic" rings could impede the large-scale
magnetic buoyancy mixing, but we do realize that they may create a real problem for the mechanism proposed by
BWNC that has been revised in our paper. Indeed, although BWNC did not discuss a formation of magnetic rings, it is difficult to understand why
the rings can not be formed at $r > r_{\rm mix}$ as easily as they are created at $r=r_{\rm mix}$.

\section{Discussion}
\label{sec:disc}

Our approximate analysis of the magnetic flux ring formation and buoyant rise in the differentially rotating radiative zone of
the bump-luminosity RGB star gives some support to a combined ``magneto-thermohaline'' mode of the RGB extra mixing, as opposed
to the pure thermohaline and pure magnetic buoyancy modes proposed by \cite{chz07a} and by BWNC, respectively.
A key component to the operation of our mixing mechanism, which is certainly present in all upper RGB and low-mass AGB stars, is
the mass inflow in the radiative zone. It makes two important things. First, the conservation of the specific angular momentum in
the mass inflow results in a steep $\Omega$-profile with a rotational shear $q\approx O(1)$ that does not appear to be strongly reduced
by the meridional circulation and turbulent diffusion (\citealt{dt00,pea06}). Second, the mass inflow entrains a diffusive
magnetic field that is probably generated via an $\alpha$\,--\,$\Omega$ or $\alpha^2$ dynamo 
just beneath the bottom of convective envelope (\citealt{nea07,nb07,m01}).
A very high ratio of the ohmic dissipation time
to the average inflow time $((\Delta r)^2/\langle\eta\rangle)/(\Delta r/\langle |\dot{r}|\rangle ) \approx 1.2\times 10^4$
guarantees that this field will be redistributed over the whole radiative zone. Because the volume occupied by this field 
is squeezed when the flow approaches the H burning shell, the field may become stronger at a smaller radius. 
In general, this random field has an unstable configuration that will decay on a short Alfv\'{e}n time scale (\citealt{s99}).
However, in their 3D MHD simulations \cite{bn06} have shown that such unstable random field 
may evolve into a stable ``twisted torus'' configuration with toroidal and poloidal field components of comparable strength. 
The poloidal field lines get wrapped around an axisymmetric torus, thus forming an approximate dipole. 
These simulations have provided the first plausible explanation of the origin and stability of dipole magnetic fields in
presently non-convective stars. It is assumed that the seed random magnetic field in those stars had been a dynamo left over from
the period of their protostellar convective contraction. 

Our RGB extra mixing model has much in common with the solar magnetic spindown model that postulates
the presence of a weak poloidal field in the solar radiative core (\citealt{mw87,chmg93}).
Note that even the values of $r_{\rm bce} = 0.996\,R_\odot$ and $\Omega (r_{\rm bce}) = 10^{-6}$\,rad\,s$^{-1}$ 
used in our RGB model are close to the corresponding solar values.
The important difference is the assumption of strong differential rotation in the radiative zone of our model,
while the present-day Sun is known to be a nearly solid-body rotator, at least above $r=0.2\,R_\odot$ (\citealt{cea03}). However,
the young Sun did possess a strong differential rotation in the core which had resulted from its spinning up during
the pre-MS contraction and angular momentum loss from the surface via a magnetized stellar wind. The magnetic spindown
model assumes that the differential rotation in the young Sun was broken by the back reaction of the Lorentz force
that emerged when the differential rotation was winding up a toroidal field from the pre-existing poloidal field (\citealt{chmg93}),
or by the magneto-rotational instability (\citealt{mlm06}).

Whereas an important role in damping large-scale toroidal field oscillations in the young Sun is thought to be played by the phase mixing (\citealt{chmg93}),
we do not think that this is also true for our magnetic buoyancy model. Indeed, the phase mixing time scale at $r=r_{\rm mix}$
in our RGB stellar model is
\bea
\label{eq:tp}
t_{\rm p} = \left(\frac{3\pi^3r^2}{\eta\omega_{{\rm A},r}^2q_{\rm A}^2}\right)^{1/3}\approx 2.9\times 10^5\ \ \mbox{yrs},
\eea
where we have used an estimate of $t_{\rm p}$ obtained by \cite{s99}, assuming that $q_{\rm A} = 1$.
This is comparable with the time needed for the whole radiative zone to be thoroughly mixed,
$t_{\rm mix} = (M_{\rm bce} - M_{\rm mix})/\dot{M}_{\rm mix}\approx 5.5\times 10^5$ yrs. Hence, long before
the toroidal field oscillations on neighboring magnetic surfaces get out of phase the magnetic buoyancy instability
will come into play, the magnetic flux rings will be formed and reach the convective envelope.

Compared with the pure thermohaline mixing, our model has the following advantages.
First, the horizontal turbulent diffusion is unlikely to hinder the buoyant rise of magnetic flux rings
with a reduced $\mu$ because the frozen-in toroidal field will not allow turbulence to penetrate plasma in the rings
and decrease the $\mu$ contrast between the ring and surrounding material. On the contrary,
there is nothing to prevent the horizontal turbulent diffusion from eroding the $\mu$ contrast in
thermohaline convective elements (\sect{sec:suppress}). 

Second, thermohaline convection
may fail to explain the operation of enhanced extra mixing in rapidly rotating Li-rich K-giants, which is needed
to activate the Cameron-Fowler mechanism, because one would expect
that the growth of ``salt fingers'' is impeded by rotation (\citealt{c99}). Oppositely, it would be natural to suppose
that magnetic flux rings are formed more efficiently in the more rapidly rotating stars. This hypothesis is
supported by the fact that the fastest rotators among young cluster solar-type stars appear to have the shortest time scale of rotational
coupling between the core and envelope (e.g., \citealt{iea07}). 

Third, whereas the efficiency of thermohaline convection in upper RGB stars is dependent exclusively on the abundance of $^3$He
left in the radiative zone, magnetic buoyancy can in principle be driven by differential rotation alone, 
provided that it succeeds in winding up a sufficiently strong toroidal field. By the end of
the RGB evolution of a low-mass star, its envelope $^3$He abundance gets depleted by a factor of ten or more (\citealt{chz07a}). 
So, what will then drive extra mixing in this star on the AGB? The higher rate of heat exchange between ``salt fingers''          
and their surrounding medium increased in proportion to the luminosity will not help because the evolutionary time scale
decreases inversely proportional to the luminosity which cancels the former effect. Thus, there is a need for an additional
driving parameter that would not let extra mixing die out. Such parameter might be the rotational velocity.
Indeed, it is known that a surprisingly fast rotation has somehow survived in red horizontal branch stars, in spite of the RGB mass loss (\citealt{p83,sp00,b03}).
It could be used on the subsequent AGB evolutionary phase to drive our magneto-thermohaline mixing.
The much lower (compared to the RGB phase) $^3$He abundance left in these stars could be compensated by a stronger toroidal magnetic field
or a larger number of magnetic flux rings generated in the presence of a higher mass inflow rate ($\dot{M}_{\rm mix}\approx 10^{-6}\,M_\odot$\,yr$^{-1}$,
according to BWNC).

Fourth, thermohaline convection cannot penetrate below the radius $r_{\rm min}\approx 0.063\,R_\odot$ at which $\mu$ has a minimum 
(the right vertical dotted line in Fig.~\ref{fig:f1}).
However, mixing down to this depth would be too shallow to reduce the surface carbon abundance (Fig.~\ref{fig:f1}a),
as required by observations (\citealt{grea00,sm03}).
An overshooting on a length scale of order $H_P$ could solve the problem (the left vertical dotted line
in Fig.~\ref{fig:f1} is placed at a distance $H_P$ below $r_{\rm min}$) but then
thermohaline ``fingers'' would have to penetrate a region of higher $\mu$ where they
experience a strong breaking. It should also be noted that the penetration of
a region with the positive $\nabla_\mu$ below $r_{\rm min}$ would reduce the average mixing rate $D_{\rm thc}\propto |\nabla_\mu|$ by
decreasing the slope of the negative $\nabla_\mu$ in the mixed radiative zone.
In the magnetic buoyancy model, there is at least a potential possibility to dredge up the nuclear processed material from below
$r_{\rm min}$. This can be done by rings with a frozen-in magnetic field of a few MG if such
are formed closer to the major H burning shell.
Of course, all the above assumptions need to be verified by more rigorous models.

Fifth, it is interesting to note that the operation of our RGB mixing mechanism seems to produce environment in which
the functioning of the $^3$He-driven thermohaline convection is impossible. Indeed, \cite{chz07b} have estimated that 
a toroidal magnetic field of $B_\varphi\approx 100$ kG would entirely suppress the thermohaline mixing at $r\approx r_{\rm mix}$,
while a field of 10 kG would be sufficient to inhibit the thermohaline instability in the upper one third of the narrow $\mu$-depression domain.
These fields are weaker than those needed for our mechanism to work.

It is also important to note that the flux rings considered in our paper are a convenient proxy for the more 
realistic $\Omega$-shaped loops that are the likely products of the magnetic buoyancy instability that operates
in a layer of strong toroidal field (e.g., \citealt{cea95}). Downflows/upflows that take place in the
loop legs during formation and rise may affect the efficiency of transport of processed material relative to the behavior
obtained in the case of a flux ring.  It is not clear how the strong rotational shear that is present within the
radiative zone would affect this process.  In principle, the only place where significant shear is required is within
the layers where $\mu$ is depressed and flux tube formation takes place.  If the overlying portion of the radiative
zone were in a state of near-uniform rotation, the loop legs might form a kind of conduit connecting the
region containing processed material with the bottom of the convection zone.  In any case, it is difficult to give
an air-tight argument for what the rotational state of the radiative zone should be, either differential rotation
of the kind assumed in our paper or near-uniform rotation enforced by the fields, that permeate the region.  

Another aspect of our computations that should be noted is that the results of sections 6.1 and 6.2 were obtained for rings
initially in thermal equilbrium with their surroundings, a state of maximal initial buoyancy.  If, alternatively,
the rings were in an initial mechanical equilibrium state (zero net force), the rise would be slower by virtue of the
heating required to overcome the neutral buoyancy at $t=0$. This would also increase the total ring number.

\section{Conclusion}

In this work, we have presented a simple model of the formation and buoyant rise of magnetic flux rings in
the radiative zone of the bump luminosity RGB star. Our model is based on ideas and equations published
by \cite{svb82}, \cite{sr83}, \cite{mw87}, \cite{chg87}, \cite{chmg93}, \cite{s99}, \cite{eea06},
BWNC, and \cite{dp08b}. It qualitatively describes a possible
mechanism for the RGB extra mixing, which we call the magneto-thermohaline mixing, as an alternative to
the pure $^3$He-driven thermohaline convection that has recently been proposed by \cite{chz07a}. For our mechanism to work, the radiative zone has to possess
a strong differential rotation and a poloidal magnetic field $B_{\rm p}\ga 1$\,--\,10 G. We assume that
the differential rotation stretches the poloidal field around the rotation axis, thus creating a strong
toroidal magnetic field $B_\varphi\approx 0.1$\,--\,1 MG. When the latter exceeds a critical value,
the buoyancy-related undular instability comes into play to form magnetic flux rings. These rings
turn out to be buoyant, therefore they rise toward the bottom of convective envelope. 

We have shown that,
when the radiative heat exchange between the ring and its surrounding medium is taken into account,
the ring's buoyant rising time increases by about five orders of magnitude compared to the case considered by BWNC, when
the ring and its surrounding medium are assumed to be in thermal equilibrium all the time. 
However, given that our model still neglects possible internal heating of the ring's material
by residual nuclear reactions and anisotropic thermal exchanges in the presence of strong oriented
magnetic field, while it uses the aerodynamic drag coefficient that is 20 times as large as the one
employed by BWNC, it is fair to say that BWNC might have fixed a safe upper limit while our paper
fixes a conservative lower limit for the magnetic ring's rising velocity.

We have found that the number of rings
needed to be present in the radiative zone at the same time to produce the observationally constrained
rate of the RGB extra mixing is unrealistically large unless these rings originate from the region of the $\mu$ inversion
maintained by the $^3$He burning. Such rings have a deficit of the mean molecular weight compared to
the bulk of the radiative zone through which they move. Their buoyancy is mainly caused by the difference
in $\mu$ rather than by a deficit in density due to the excess magnetic pressure. The frozen-in toroidal
magnetic field is still needed for the rings to remain cohesive while rising. That is why we have coined
the term ``magneto-thermohaline'' mixing. Our model has some advantages over the pure
thermohaline mixing model, the most important of which being
the robustness of the magnetic rings against the eroding effect produced by the horizontal turbulent diffusion. 
Leaving aside the problem of the "parasitic" rings that are formed at $r\ga r_{\rm mix} + 0.05\,R_\odot$, our model looks promising. 
However, because it is based
on a number of assumptions whose legality is impossible to confirm in the framework of our 1D computations
we call for its future verification by 3D MHD simulations.

\acknowledgements
PAD and MP acknowledge support from the NASA grant NNG05 GG20G.
The National Center for Atmospheric Research is sponsored by the National Science Foundation.


\clearpage
\begin{deluxetable}{llllll}
\tablecolumns{6}
\tabletypesize{\footnotesize}
\tablecaption{RGB Model Structure Parameters}
\tablewidth{0pt}
\tablehead{
\colhead{} & \colhead{} & \multicolumn{2}{c}{Our Model} &
\multicolumn{2}{c}{BWNC Model} \\
\cline{3-4} \cline{5-6} \\
\colhead{Parameter} & \colhead{Units} &
          \colhead{$r_{\rm mix}=0.0500\,R_\odot$} & \colhead{$r_{\rm bce}=0.996\,R_\odot$} &
          \colhead{$r_{\rm mix}=0.0495\,R_\odot$} & \colhead{$r_{\rm bce}=0.912\,R_\odot$} }
\startdata
$M_r$ & $M_\odot$ & 0.309 & 0.331 & 0.252 & 0.256 \\
$H_P$ & $r$ & 0.257 & 0.449 & 0.371 & 0.622 \\
$\mu$ & AMU & 0.593406 & 0.593474 & & \\
$T$ & K & $2.16\times 10^7$ & $2.02\times 10^6$ & $2.51\times 10^7$ & $2.26\times 10^6$ \\
$\rho$ & g\,cm$^{-3}$ & 14.8 & $1.03\times 10^{-2}$ & 5.19 & $2.55\times 10^{-3}$ \\
$P$ & dyn\,cm$^{-2}$ & $4.49\times 10^{16}$ & $2.93\times 10^{12}$ & $1.87\times 10^{16}$ & $8.49\times 10^{11}$ \\
$K$ & cm$^2$\,s$^{-1}$ & $9.35\times 10^7$ & $8.15\times 10^{10}$ & & \\
$N_T^2$ & s$^{-2}$ & $5.40\times 10^{-4}$ & $1.13\times 10^{-9}$ & & \\
$N_\mu^2$ & s$^{-2}$ & $1.48\times 10^{-7}$ & $0.00$ & & \\
$\nu_{\rm mol}$ & cm$^2$\,s$^{-1}$ & $1.80\times 10^2$ & $6.76\times 10^2$ & & \\
$\nu_{\rm rad}$ & cm$^2$\,s$^{-1}$ & $1.74\times 10^2$ & $1.43\times 10^4$ & & \\
$\eta$ & cm$^2$\,s$^{-1}$ & 19.3 & $6.90\times 10^2$ & & \\
$\kappa$ & cm$^2$\,g$^{-1}$ & 0.382 & 0.728 & & \\
$|\dot{r}|$ & cm\,s$^{-1}$ & $3.91\times 10^{-5}$ & $8.50\times 10^{-5}$ & & \\
$\Omega$ & rad\,s$^{-1}$ & $10^{-3}$ & $10^{-6}$ & & \\
\enddata
   \label{tab:tab1}
\end{deluxetable}


\clearpage
\begin{deluxetable}{llllll}
\tablecolumns{6}
\tabletypesize{\footnotesize}
\tablecaption{Magnetic Flux Ring Parameters}
\tablewidth{0pt}
\tablehead{
\colhead{} & \colhead{} & \multicolumn{2}{c}{Our Model\ \ ($\Delta\mu = 5\times 10^{-5}$)} &
\multicolumn{2}{c}{BWNC Model\ \ ($\Delta\mu = 0$)} \\
\cline{3-4} \cline{5-6} \\
\colhead{Parameter} & \colhead{Units} &
          \colhead{$B_r = 7.3$\,G} & \colhead{$B_r = 73$\,G} & 
          \colhead{RGB-1} & \colhead{RGB-2} }
\startdata
$a_0$ & cm & $1.3\times 10^7$ & $9.7\times 10^6$ & $6.5\times 10^6$ & $1.5\times 10^7$ \\
$a_0$ & $H_P$ & $1.5\times 10^{-2}$ & $1.1\times 10^{-2}$ & $5.1\times 10^{-3}$ & $1.2\times 10^{-2}$ \\
$t_{\rm b}$ & yr & $2.9\times 10^2$ & $1.6\times 10^2$  & 0.19 & 1.1 \\
$\langle v_{\rm b}\rangle$ & cm\,s$^{-1}$ & 7.3 & 13 & $9.8\times 10^3$ & $1.8\times 10^3$ \\
$(B_\varphi)_0$ & kG & 254  & 254 & 380 & 48 \\
$\beta_0$ & & $1.8\times 10^7$ & $1.8\times 10^7$ & $3.3\times 10^6$ & $2.0\times 10^8$ \\
$(B_\varphi)_{\rm bce}$ & kG & 3.5 & 3.5 & 3.5 & 0.44 \\
$N$ & & $1.3\times 10^2$ & $1.3\times 10^2$ & 1.0 & 1.0 \\
$\dot{N}=N/t_{\rm b}$ & yr$^{-1}$ & 0.45 & 0.83 & 5.3 & 0.91 \\
\enddata
   \label{tab:tab2}
\end{deluxetable}


\begin{thebibliography}{}

\bibitem[Acheson(1978)]{a78}
Acheson, D.~J.~1978, Phil. Trans. R. Soc. Lond. A, 289, 459

\bibitem[Aoki et al.(2008)]{aea08}
Aoki, W., Beers, T.~C., Sivarani, T., Marsteller, B., Lee, Y.~S., Honda, S.,
Norris, J.~E., Ryan, S.~G., \& Carollo, D.~2008, ApJ, 678, 1351

\bibitem[Behr et al.(2003)]{b03}
Behr, B.~B.~2003, ApJS, 149, 101

\bibitem[Blackman et al.(2001)]{bea01}
Blackman, E.~G., Frank, A., Markiel, J.~A., Thomas, J.~H., \& Van Horn, H.~M.~2001, Nature, 409, 485

\bibitem[Braithwaite \& Nordlund(2006)]{bn06}
Braithwaite, J., \& Nordlund, \AA.~2006, A\&A, 450, 1077

\bibitem[Busso et al.(2007)]{bea07}
Busso, M., Wasserburg, G.~J., Nollett, K.~M., \& Calandra, A.~2007, ApJ, 671, 802 (BWNC)

\bibitem[Caligari, Moreno-Insertis, \& Schussler(1995)]{cea95}
Caligari, P., Moreno-Insertis, F., \& Schussler, M.~1995, ApJ, 441, 886

\bibitem[Cameron \& Fowler(1971)]{cf71}
Cameron, A.~G.~W., \& Fowler, W.~A.~1971, ApJ, 164, 111

\bibitem[Canuto(1999)]{c99}
Canuto, V.~M.~1999, ApJ, 524, 311

\bibitem[Chaboyer \& Zahn(1992)]{chz92}
Chaboyer, B., \& Zahn, J.-P.~1992, A\&A, 253, 173

\bibitem[Chanam\'{e}, Pinsonneault, \& Terndrup(2005)]{chea05}
Chanam\'{e}, J., Pinsonneault, M., \& Terndrup, D.~M.~2005, ApJ, 631, 540

\bibitem[Charbonneau \& MacGregor(1993)]{chmg93}
Charbonneau, P., \& MacGregor, K.~B.~1993, ApJ, 417, 762

\bibitem[Charbonnel \& Do Nascimento(1998)]{chdn98}
Charbonnel, C., \& Do Nascimento, J.~D., Jr.~1998, A\&A, 336, 915

\bibitem[Charbonnel, Brown, \& Wallerstein(1998)]{chea98}
Charbonnel, C., Brown, J.~A., \& Wallerstein, G.~1998, A\&A, 332, 204

\bibitem[Charbonnel \& Balachandran(2000)]{chb00}
Charbonnel, C., \& Balachandran, S.~C.~2000, A\&A, 359, 563

\bibitem[Charbonnel \& Zahn(2007a)]{chz07a}
Charbonnel, C., \& Zahn, J.-P.~2007a, A\&A, 467, L15

\bibitem[Charbonnel \& Zahn(2007b)]{chz07b}
Charbonnel, C., \& Zahn, J.-P.~2007b, A\&A, 476, L29

\bibitem[Choudhuri \& Gilman(1987)]{chg87}
Choudhuri, A.~R., \& Gilman, P.~A.~1987, ApJ, 316, 788

\bibitem[Couvidat et al.(2003)]{cea03}
Couvidat, S., Garc\'{\i}a, R.~A., Turck-Chi\`{e}ze, Corbard, T., Henney, C.~J.,
\& Jim\'{e}nez-Reyes, S.~2003, ApJ, 597, L77

\bibitem[Denissenkov \& Tout(2000)]{dt00}
Denissenkov, P.~A., \& Tout, C.~A.~2000, MNRAS, 316, 395

\bibitem[Denissenkov \& VandenBerg(2003)]{dv03}
Denissenkov, P.~A., \& VandenBerg, D.~A.~2003, ApJ, 593, 509

\bibitem[Denissenkov \& Herwig(2004)]{dh04}
Denissenkov, P.~A., \& Herwig, F.~2004, ApJ, 612, 1081

\bibitem[Denissenkov et al.(2006)]{dea06}
Denissenkov, P.~A., Chaboyer, B., \& Li, K.~2006, ApJ, 641, 1087

\bibitem[Denissenkov \& Pinsonneault(2008a)]{dp08a}
Denissenkov, P.~A., \& Pinsonneault, M.~2008a, ApJ 679, 1541

\bibitem[Denissenkov \& Pinsonneault(2008b)]{dp08b}
Denissenkov, P.~A., \& Pinsonneault, M.~2008b, ApJ, 684, 626

\bibitem[Denissenkov \& Weiss(1996)]{dw96}
Denissenkov, P.~A., \& Weiss, A.~1996, A\&A, 308, 773

\bibitem[Drake et al.(2002)]{dea02}
Drake, N.~A., de la Reza, R., da Silva, L., \& Lambert, D.~L.~2002, AJ,
123, 2703

\bibitem[Eggleton, Dearborn, \& Lattanzio(2006)]{eea06}
Eggleton P.~P., Dearborn, D.~S.~P., \& Lattanzio, J.~C.~2006, Science, 314, 1580

\bibitem[Fan(2001)]{f01} 
Fan, Y.~2001, ApJ, 546, 509

\bibitem[Gratton et al.(2000)]{grea00}
Gratton, R.~G., Sneden, C., Carretta, E., \& Bragaglia, A.~2000, A\&A, 354, 169

\bibitem[Irwin et al.(2007)]{iea07}
Irwin, J., Hodgkin, S., Aigrain, S., Hebb, L., Bouvier, J., Clarke, C.,
Moraux, E., \& Bramich, D.~M. 2007, MNRAS, 377, 741

\bibitem[Kippenhahn et al.(1980)]{kea80}
Kippenhahn, R., Ruschenplatt, G., \& Thomas, H.-C.~1980, A\&A, 91, 175

\bibitem[Lebzelter et al.(2008)]{lea08}
Lebzelter, T., Lederer, M.~T., Cristallo, S., Hinkle, K.~H., Straniero, O., \& Aringer, B.~2008,
A\&A, 486, 511L

\bibitem[MacDonald \& Mullan(2004)]{mdm04}
MacDonald, J., \& Mullan, D.~J.~2004, MNRAS, 348, 702

\bibitem[MacGregor \& Cassinelli(2003)]{mgc03}
MacGregor, K.~B., \& Cassinelli, J.~P.~2003, ApJ, 586, 480

\bibitem[Maeder(2003)]{m03}
Maeder, A.~2003, A\&A, 399, 263 

\bibitem[Masseron et al.(2006)]{mea06}
Masseron, T., Van Eck, S., Famaey, B., Goriely, S., Plez, B., Siess, L.,
Beers, T.~C., Primas, F., \& Jorissen, A.~2006, A\&A, 455, 1059

\bibitem[Menou \& Le Mer(2006)]{mlm06}
Menou, K., \& Le Mer, J.~2006, ApJ, 650, 1208

\bibitem[Mestel(2001)]{m01}
Mestel, L.~2001, in Magnetic Fields across the Hertzsprung-Russel Diagram, eds. G. Mathys,
S. K. Solanki, and D. T. Wickramasinghe, ASP Conference Series, 248, 3

\bibitem[Mestel \& Weiss(1987)]{mw87}
Mestel, L., \& Weiss, N.~O.~1987, MNRAS, 226, 123

\bibitem[Moss(2003)]{moss03}
Moss, D.~2003, A\&A, 403, 693 

\bibitem[Newsham \& Terndrup(2007)]{nt07}
Newsham, G. \& Terndrup, D.~M.~2007, ApJ, 664, 332 

\bibitem[Nollett, Busso, \& Wasserburg(2003)]{nbw03}
Nollett, K.-M., Busso, M., \& Wasserburg, G.~J.~2003, ApJ, 582, 1036

\bibitem[Nordhaus, Blackman, \& Frank(2007)]{nea07}
Nordhaus, J., Blackman, E.~G., \& Frank, A.~2007, MNRAS, 376, 599 

\bibitem[Nordhaus \& Blackman(2007)]{nb07}
Nordhaus, J., \& Blackman, E.~G.~2008, in IXth Torino Workshop on Evolution and Nucleosynthesis
in AGB Stars and the IInd Perugia Workshop on Nuclear Astrophysics, AIP Conference Proc., 1001, 306

\bibitem[Palacios et al.(2003)]{pea03}
Palacios, A., Charbonnel, C., Talon, S., \& Forestini, M.~2003, A\&A, 399, 603

\bibitem[Palacios et al.(2006)]{pea06}
Palacios, A., Charbonnel, C., Talon, S., \& Siess, L.~2006, A\&A, 453, 261

\bibitem[Palacios \& Brun(2006)]{pb06}
Palacios, A., \& Brun, A.~S.~2006, in IAU Symp. 239, Convection in Astrophysics,
ed. F. Kupka, I. Roxburgh \& K.~L. Chan, 431 (arXiv:astro-ph/0610040v1)

\bibitem[Peterson(1983)]{p83}
Peterson, R.~C.~1983, ApJ, 275, 737

\bibitem[Pitts \& Tayler(1985)]{pt85}
Pitts, E., \& Tayler, R.~J.~1985, MNRAS, 216, 139

\bibitem[Ryan et al.(2005)]{rea05}
Ryan, S.~G., Aoki, W., Norris, J.~E., \& Beers, T.~C.~2005, ApJ, 635, 349

\bibitem[Schmitt \& Rosner(1983)]{sr83}
Schmitt, J.~H.~M.~M., \& Rosner, R.~1983, ApJ, 265, 901

\bibitem[Siess, Dufour, \& Forestini(2000)]{sea00}
Siess, L., Dufour, E., \& Forestini, M.~2000, A\&A, 358, 593

\bibitem[Sills \& Pinsonneault(2000)]{sp00}
Sills, A., \& Pinsonneault, M.~H.~2000, ApJ, 540, 489

\bibitem[Sivarani et al.(2006)]{sea06}
Sivarani, T., et al.~2006, A\&A, 459, 125

\bibitem[Smith \& Martell(2003)]{sm03}
Smith, G.~H., \& Martell, S.~L.~2003, PASP, 115, 1211

\bibitem[Spruit(1999)]{s99}
Spruit, H.~C.~1999, A\&A, 349, 189

\bibitem[Spruit \& van Ballegooijen(1982)]{svb82}
Spruit, H.~C., \& van Ballegooijen, A.~A.~1982, A\&A, 106, 58 

\bibitem[Stern(1960)]{s60}
Stern, M.~E.~1960, Tellus, 12, 172

\bibitem[Sweigart \& Mengel(1979)]{sm79}
Sweigart, A.~V., \& Mengel, J.~G.~1979, ApJ, 229, 624

\bibitem[Talon \& Zahn(1997)]{tz97}
Talon, S., \& Zahn, J.-P.~1997, A\&A, 317, 749

\bibitem[Talon et al.(1997)]{tea97}
Talon, S., Zahn, J.-P., Maeder, A., \& Meynet, G.~1997, A\&A, 322, 209

\bibitem[Tayler(1973)]{t73}
Tayler, R.~J.~1973, MNRAS, 161, 365

\bibitem[Ulrich(1972)]{u72}
Ulrich, R.~K.~1972, ApJ, 172, 165

\bibitem[Vauclair(2004)]{v04}
Vauclair, S.~2004, ApJ, 605, 874

\bibitem[Wasserburg et al.(2006)]{wea06}
Wasserburg, G.~J., Busso, M., Gallino, R., \& Nollett, K.~M.~2006, Nucl. Phys. A, 777, 5

\bibitem[Zahn(1992)]{z92}
Zahn, J.-P.~1992, A\&A, 256, 115

\end{thebibliography}
\end{document}